\documentclass{aastex631}
\usepackage{epstopdf}
\usepackage{natbib}
\usepackage{graphicx}
\usepackage{CJK}

\usepackage[percent]{overpic}
\usepackage{times}
\usepackage{graphicx}
\usepackage{amsmath}
\usepackage{subfigure}
\usepackage{natbib}
\usepackage{txfonts}
\usepackage{journals}
\usepackage{xcolor}
\usepackage[utf8]{inputenc}
\usepackage[overload]{empheq}
\usepackage{verbatim}

\newcommand       \Teff         {T_{\rm {eff}}}
\defcitealias{2021ApJS..254...38S}{Paper\,I}
\submitjournal{AJ}
\begin{document}
\begin{CJK*}{UTF8}{gbsn}

\title[Identification and distance measurement of dust clouds at high latitude by a clustering hierarchical algorithm]
{Identification and distance measurement of dust clouds at high latitude by a clustering hierarchical algorithm}

\author[0000-0002-2473-9948]{Mingxu Sun (孙明旭)}
\affiliation{Department of Physics,
               Hebei Key Laboratory of Photophysics Research and Application,
               Hebei Normal University,
               Shijiazhuang 050024, P.\,R.\,China}
\author[0000-0003-3168-2617]{Biwei Jiang(姜碧沩)}
\affiliation{School of Physics and Astronomy,
               Beijing Normal University,
               Beijing 100875, P.\,R.\,China}
\author[0000-0001-5737-6445]{Helong Guo(郭贺龙)}
\affiliation{South-Western Institute for Astronomy Research,
               Yunnan University,
               Kunming 650500, P.\,R.\,China}
\author[0000-0003-1359-9908]{Wenyuan Cui(崔文元)}
\affiliation{Department of Physics,
               Hebei Key Laboratory of Photophysics Research and Application,
               Hebei Normal University,
               Shijiazhuang 050024, P.\,R.\,China}
               
\correspondingauthor{Mingxu Sun}
\email{mxsun@hebtu.edu.cn}

\date{Accepted 12-Aug-2024. Received 04-Jan-2024; in original form 27-Dec-2023}

\label{firstpage}

\begin{abstract}

We present a catalog of dust clouds at high Galactic latitude based on the Planck 857 GHz dust emission data. Using a clustering hierarchical algorithm, 315 dust cloud at high Galactic latitudes are identified. Additionally, using the optical and ultraviolet extinction of 4 million and 1 million stars, respectively, provided by Sun et al., we derive the distances and physical properties for 190 high Galactic latitude dust clouds and the ultraviolet excess ratios for 165 of them. Through the study of color excess ratios, this work confirms that molecular clouds with  large Galactic distances and low extinction likely have a higher proportion of small-sized dust grains. In addition, clouds with well-defined distances in the catalog are used to trace the local bubble, showing good consistency with the boundary of the local bubble from the literature.

\end{abstract}


\section{Introduction} \label{sec:intro}

The identification and study of dust clouds in the Milky Way is critical to our understanding of the initial mass function of stars and the formation of galaxies \citep{1999ASIC..540....3B}. In the early days, the identification of dust clouds relied on detecting dark areas in survey photographs \citep{1919ApJ....49....1B,1927cdos.book.....B}. Later in a similar way, \citet{2005PASJ...57S...1D} identified 2448 dark clouds in their two-dimensional extinction map of the Galactic plane, which was later upgraded by \citet{2011PASJ...63S...1D} to identify 7614 clouds in the entire sky. \citet{2020PASP..132e4301P} utilize open astronomical source, image processing, and neural network libraries to apply this method to various GLIMPSE and other Galactic plane data, creating a catalog of Infrared Dark Clouds. In addition, the CO molecule is also a good tracer of dust clouds, as CO is typically the most easily observable molecule and the molecules are well mixed with dust grains. Many large-scale CO survey projects have discovered numerous molecular and dust clouds in the Milky Way \citep[e.g.,][]{1985ApJ...295..402M,1997A&A...327..325M,2001ApJ...547..792D}. 
A multitude of research endeavors focusing on the identification and study of molecular clouds have emerged from these large-scale CO survey projects.\citep[e.g.,][]{2003ApJ...583..280B,2010ApJ...723..492R,2015ARA&A..53..583H,2016ApJ...822...52R,2017ApJ...834...57M,2021A&A...654A.144B}.

High-latitude clouds typically exhibit low extinction (translucent) states \citep{1991ApJ...366..141V,2002A&A...383..631D}, where $A_{\rm V}$ is typically around 1-2 \citep{2006ARA&A..44..367S}. Large-scale diffuse emission was first discovered by blue plates and H-alpha plates in the Milky Way \citep{1955Obs....75..129D}. \citet{1976AJ.....81..954S} published the first detailed optical image of high-latitude clouds in the Galactic polar cap survey. One unexpected result from the Infrared Astronomical Satellite (IRAS) was the discovery of large-scale, extended filamentary emission at 60 and 100 $\mu$m, which \citet{1984ApJ...278L..19L} called "infrared cirrus". Coincident with the discovery of the IRAS cirrus, \citet[BMM,][]{1984ApJ...282L...9B} found a large population of previously uncataloged high-latitude ($|b|>25\degr$) molecular clouds. These molecular cloud maps \citep[MBM,][]{1985ApJ...295..402M} displayed a variety of morphologies, sometimes spanning over 10$\degr$. Based on the large-scale CO emission, BMM suggested that the IRAS cirrus may be associated with the radiation from high-latitude molecular clouds. \citet{1986ApJ...306L.101W} compared the MBM and IRAS data and found that in all cases, these high-latitude molecular clouds are associated with the 100 $\mu$m cirrus emission from IRAS. \citet{1986ApJ...304..466K} also studied high-latitude (including some low-latitude) clouds in CO. Besides, \citet{1988ApJ...334..815D} used IRAS 100 $\mu$m data in conjunction with the Berkeley atomic HI survey \citep{1974A&AS...14....1H} to detect 516 infrared excess clouds (IRECs) with $|b|>5\degr$. \citet{1998ApJ...507..507R} combined the DIRBE/COBE and Leiden-Dwingeloo H I surveys \citep{1997agnh.book.....H} to extract 60 previously identified clouds from infrared spectra and discovered 81 new clouds.\citet{2002A&A...383..631D} cross-identified 21 dust cloud catalogues in the Galaxy, and the unified catalogue includes 525 high-latitude clouds.

Determining the distance to clouds is important, but can often pose challenges. Estimating the distance of clouds by the Galactic kinematics \citep[e.g.][]{2009ApJ...699.1153R} or associating the cloud with objects of known distance \citep[e.g.][]{2008AN....329...10M} can be applied to the cases in the Galactic plane. However, the two methods may not be suitable for high-latitude clouds due to their significant deviation from the rotation curve of the Galactic disk and the absence of young massive stars within them. The presence of dense dust within molecular clouds allows for the estimation of their distance by measuring the extinction toward stars at known distances \citep{2009ApJ...692...91G,2017MNRAS.472.3924C}. By comparing the density of stars located in front of molecular clouds with minimal extinction to predictions of Galactic models, the distances of many clouds are estimated by \citet{2009ApJ...703...52L} and \citet{2011A&A...535A..16L}. Using the multi-band photometry data acquired from PanSTARRS-1 \citep{2010SPIE.7733E..0EK} and the resulting color indices of  a large number of stars, \citet{2014ApJ...786...29S} determined the distances to 18 prominent star-forming regions and 108 high Galactic latitude molecular clouds from the extinction breakpoint. 

The \emph{Gaia}/DR2 catalog \citep[\emph{Gaia}/DR2;][]{2018A&A...616A...1G} revolutionizes the determination of distances to dust clouds by providing distance measurements to over a billion stars. Using the \emph{Gaia}/DR2 data, \citet{2019ApJ...879..125Z} present a comprehensive catalog of accurate distances to local dust clouds based on the breakpoint of the extinction. \citet{2019A&A...624A...6Y} derive distances to dust clouds at high Galactic latitudes ($|b| > 10\degr$) using parallax measurements \citep{2018A&A...616A...2L} and G-band extinction ($A_{\rm G}$) measurements from \emph{Gaia}/DR2. By using a three-dimensional dust reddening map and estimating color excesses and distances for more than 32 million stars, \citet{2020MNRAS.493..351C} identify 567 clouds within 4 kpc of the Sun at low Galactic latitudes ($|b|\le10\degr$) using a hierarchical structure identification method and a dust model fitting algorithm for distance estimation. In the southern sky, \citet{2022MNRAS.511.2302G} identify 250 clouds with accurate distance estimates. These studies benefit from the large number of stars within individual dust clouds and the reliable distance estimates provided by \emph{Gaia}/DR2, with typical distance errors of only about 5 percent.

With much smaller extinction than those in the disk, high-latitude clouds need a precise measurement of extinction and distance to the stars for both identification and distance determination. \citet{2021ApJS..254...38S} used a combination of photometric and spectroscopic data to determine the color excess $E_{\rm G_{BP},G_{RP}}$ and $E_{\rm NUV,G_{BP}}$ which are typically accurate to 0.01 mag and 0.1 mag respectively and better than that derived from photometry only. With the accurate color excess and stellar distance from \emph{Gaia}/EDR3, \cite{2021ApJS..256...46S} determined the distances to 66 \citet{1985ApJ...295..402M} (MBM) molecular clouds. 

This work is a continuous of the previous one but extend to a large area at high latitude.  Moreover, this work will use a clustering hierarchical algorithm to identify high latitude dust clouds from the Planck 857 GHz map \citep{2020A&A...641A...1P} for which the distance, size, extinction, and color excess will be calculated.

\section{The Data}\label{sec:dat}

The Planck 857 GHz map \citep{2020A&A...641A...1P} is used for dust cloud identification.  Planck is the first submillimeter mission to map the entire sky to sub-Jansky sensitivity with angular resolution better than 10$\arcmin$.  At 857 GHz, the uncertainty in its Galactic emission zero level is 0.0147 MJy sr$^{-1}$, CIB level is 0.64 MJy sr$^{-1}$, and the uncertainty in its CIB zero level is 0.077 MJy sr$^{-1}$ \citep{2020A&A...641A...1P}. This map has a spatial resolution of approximately $4.22\arcmin$, which is slightly better than that of the IRAS 100 $\mu$m image \citep{1998ApJ...500..525S,2005ApJS..157..302M}. 

When compared to the CO survey \citep{2001ApJ...547..792D}, the Planck 857 GHz survey provides more comprehensive coverage at high Galactic latitudes in that the CO survey primarily focused on the range of Galactic latitudes with $|b|<30\degr$, whereas the Planck 857 GHz survey covers all the high-latitude sky area.

The extinction values are taken from \citet{2021ApJS..254...38S} that provides the values of the color excesses, $E_{{\rm G_{BP},G_{RP}}}$ between the $Gaia/BP$ and $Gaia/RP$ bands and $E_{{\rm NUV,G_{BP}}}$ between the $GALEX/NUV$ and $Gaia/BP$ bands for over 4 million and 1 million dwarf stars, respectively, which are mainly located at high latitudes. Combining stellar parameters from the LAMOST \citep{2015RAA....15.1095L} and GALAH \citep{2021MNRAS.506..150B} spectroscopic survey, and photometric data ($G_{\rm BP}$, $G_{\rm RP}$ and $NUV$) from \emph{Gaia}/EDR3 \citep{2020arXiv201201533G} and GALEX \citep{2014Ap&SS.354..103B} and using the blue-edge method (c.f. \citet{2017AJ....153....5J} and \citet{2018ApJ...861..153S}), the average error of $E_{{\rm G_{BP},G_{RP}}}$ and $E_{{\rm NUV,G_{BP}}}$ is $\sim$ 0.01 mag and 0.1 mag respectively. The stars with the color excess of \citet{2021ApJS..254...38S} and distance from \emph{Gaia}/EDR3 will be the tracers for the extinction and distance of the high-latitude clouds.
 
This work estimates the gas cloud masses from HI and CO observations. For the HI observation, the HI4PI survey, which is based on the completed Effelsberg-Bonn HI Survey (EBHIS) and the third revision of the Galactic All-Sky Survey (GASS), is chosen \citep{2016A&A...594A.116H}. The HI4PI survey has an angular resolution of 16.2$'$ and a sensitivity of 43 mK. For the CO observation, the Planck survey is chosen, and the Type2 data is used \citep{2014A&A...571A..13P}. Type2 utilizes the intensity maps from multiple channels to isolate the contribution of CO from other astrophysical emission sources. The average standard deviation of the CO (J = 1$\rightarrow$0) is 1.77 $\mathrm{K_{RJ} \cdot km \cdot s^{-1}}$. The Planck data provides the CO integrated intensity in the unit of $\mathrm{K_{RJ} \cdot km \cdot s^{-1}}$. The conversion factor between the radio antenna temperature $\mathrm{T_{RJ}}$ and the thermodynamic temperature $\mathrm{T_{therm}}$ as $\mathrm{C_{T_{RJ}\rightarrow T_{therm}} \approx 1.287}$, which is applied to the Planck data unit conversion.

\section{Identification of the dust clouds}

\subsection{A clustering hierarchical algorithm}

Our method begins by using the Planck 857 GHz image to isolate individual dust clouds. To achieve this, a clustering hierarchical algorithm is used, which identifies coherent structures in the Galactic ($l, b$) position space. Specifically, the astrodendro Python program developed by \citet{2008ApJ...679.1338R} is used to generate tree representations of the hierarchical structure of nested isosurfaces in the two-dimensional 857 GHz image. The main focus of this study is on relatively diffuse dust clouds at high Galactic latitudes, which are much less prone to contamination from other clouds and rarely exhibit line-of-sight adjacency or overlap. This provides us with favorable conditions to utilize the 857 GHz image for dust cloud identification. 

\begin{figure*}
\centering
\includegraphics[width=1\textwidth,angle=0]{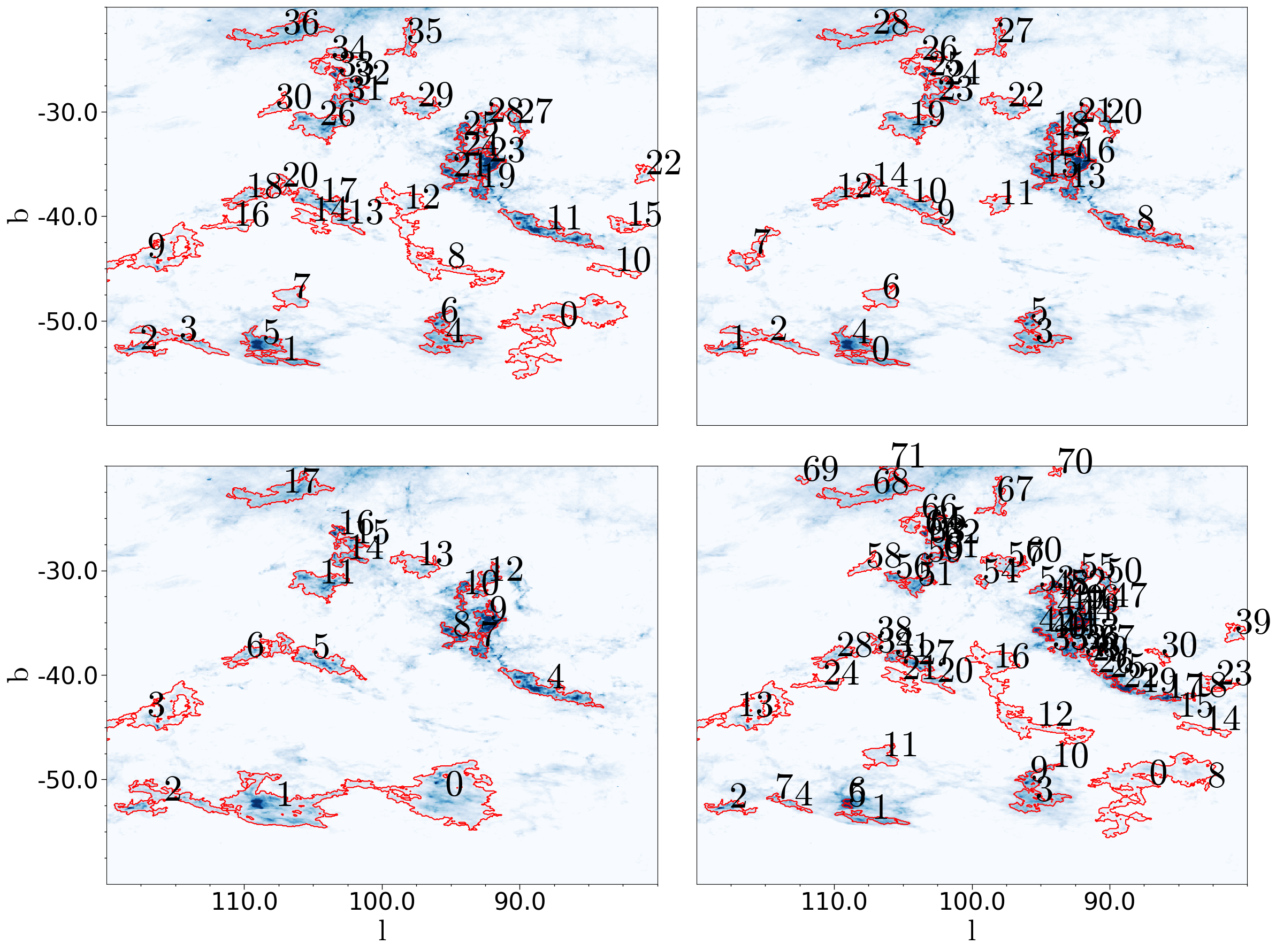}
\caption{ The cloud identification results using different parameters for hierarchical clustering within the region of $80\degr<l<120\degr$ and $20\degr<b<60\degr$. Top-left panel: min value = 1.74, min delta = 3.48, and min npix = 225; Top-right panel: min value = 3.48, min delta = 3.48, and min npix = 225; Bottom-left panel: min value = 1.74, min delta = 5.22, and min npix = 225; Bottom-right panel: min value = 1.74, min delta = 3.48, and min npix = 25. The black numbers represent the cloud numbers.}
\label{detu}
\end{figure*}

There are three input parameters in the `Dendrogram algorithm': (i) `min\_value', which masks structures peaking below a specific threshold; (ii) `min\_delta', which defines a minimum height required for a structure to exclude any local maxima identified because of the noise; and (iii) `min\_npix', which establishes the minimum number of pixels necessary for a structure to be considered a distinct entity. To demonstrate the impact of parameters on cloud identification results, Figure~\ref{detu} presents the cloud identification results with different parameters in the $80\degr<l<120\degr$ and $20\degr<b<60\degr$ region. From these images, it can be observed that increasing the  `min\_value' raises the $I_{\rm 857 GHz}$ threshold for identifying regions as clouds, resulting in the exclusion of some clouds and a reduction in the identified cloud sizes. Only denser clouds and denser regions within clouds can be identified. Increasing the  `min\_delta' raises the requirement for the contrast in $I_{\rm 857 GHz}$ between clouds and their surroundings, which may lead to the exclusion of certain clouds. Additionally, as min\_delta increases, the uppermost leaves may no longer be recognized as distinct structures. Instead, they will merge with the adjacent trunk, forming a new uppermost leaf. This results in the merging of some clouds and an expansion of the cloud regions. The `min\_npix' parameter sets the minimum size requirement for clouds. Decreasing this value leads to the identification of more small-sized clouds or the fragmentation of large clouds into small components. It is evident that these parameters have a significant impact on cloud identification results. Determining these parameters depends on the characteristics of the data and the objectives of the study.

\subsection{The dust cloud identification}
\begin{figure*}
\centering
\includegraphics[width=1\textwidth,angle=0]{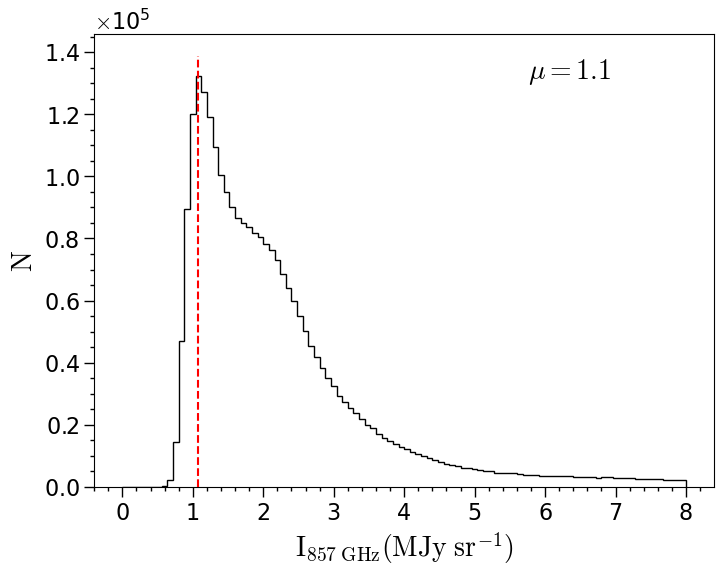}
\caption{The stellar number distribution with $I_{\rm 857GHz}$ at high Galactic latitude ($|b|>20^{\circ}$). $\rm \mu$ represent the stellar source number $I_{\rm 857GHz}$. The red dashed line indicates $I_{\rm 857GHz}$= 1.1 line.}
\label{qztf}
\end{figure*}

The Planck 857 GHz image used as input for hierarchical clustering is presented with the pixel size of $4\arcmin$. To clarify the input parameters in the `Dendrogram algorithm', we combine the optical color excess catalog from \citet{2021ApJS..254...38S} with the Planck 857 GHz image based on the consistency of sightlines, presenting the stellar source number distribution at high Galactic latitude ($|b|>20^{\circ}$) for different Planck 857 GHz intensities, as shown in Figure~\ref{qztf}. Generally, the density of stellar sources is related to extinction or Planck 857 GHz intensity. In the sightlines of the diffuse interstellar medium, the stellar source density is high, while in the sightlines of clouds, the stellar source density is low. Therefore, the peak of the stellar source number distribution for different Planck 857 GHz intensities reflects the Planck 857 GHz intensity corresponding to the diffuse interstellar medium, which can be considered the noise threshold for cloud signal detection. In this work, we set min\_value = 1.1 $\rm MJy\ sr^{-1}$ (peak stellar number density) + 0.64 $\rm MJy\ sr^{-1}$ (the CIB level) = 1.74 $\rm MJy\ sr^{-1}$ which corresponds to approximately 0.07 mag in $E_{{\rm G_{BP},G_{RP}}}$, and min\_delta = 2 $\times$ min\_values = 3.48 $\rm MJy\ sr^{-1}$ corresponding to about 0.14 mag in $E_{{\rm G_{BP},G_{RP}}}$. Since the main objective of this study is to investigate relatively diffuse clouds at high Galactic latitudes instead of dense cores, a relatively large value for the `min\_npix' is chosen to be 225, which corresponds to about 1 $\rm deg^2$ in angular size with the Planck 857 GHz image presented with the pixel size of $4\arcmin$. The resultant `Dendrogram leaves' are the regions of density enhancement, or individual dust cloud in our case.

\begin{figure*}
\centering
\includegraphics[width=1\textwidth,angle=0]{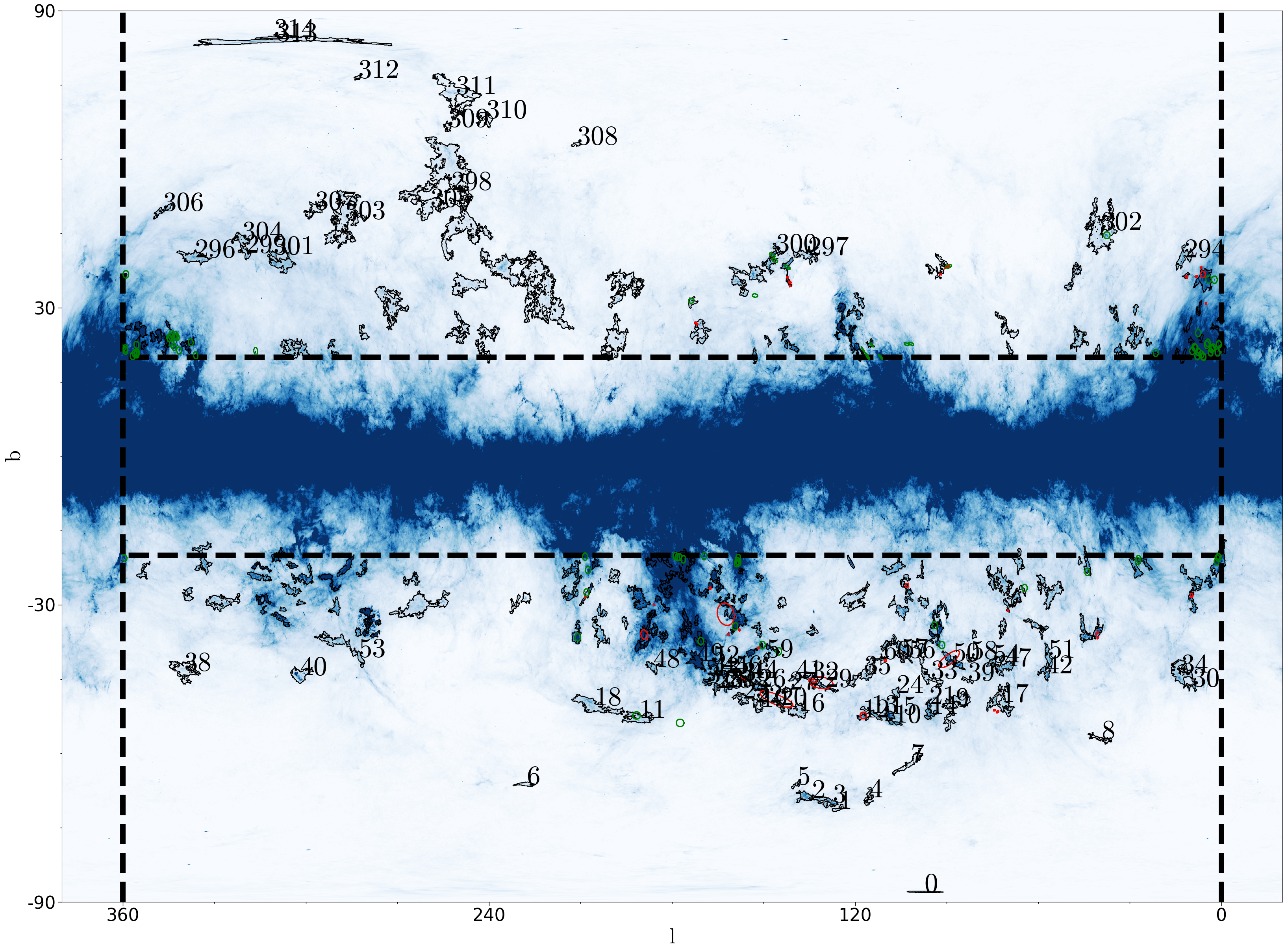}
\caption{The color map is the Planck 857GHz map. The black solid lines enclose the region containing clouds with centers within the range of $0<l<360\degr$ and $|b|>20\degr$ by a clustering hierarchical algorithm. The black numbers represent the cloud numbers with $|b|>40^{\circ}$. The red and green ellipses represent the MBM molecular clouds with and without sizes (default radius of 90$^\prime$). The dashed black lines represent the $|b|=20\degr$ lines.}
\label{qztd}
\end{figure*}

In order to avoid edge effects, the spatial coverage of the Planck 857 GHz image is extended beyond the conventional Galactic longitude range of 0$\degr$ to 360$\degr$ in that the part of 360$\pm25\degr$ is repeatedly used.  After the cloud identification, only the clouds whose central position is within the range of $0\degr<l<360\degr$ and $|b|>20\degr$ are retained. The identified clouds are displayed over the Planck 857 GHz image in Figure~\ref{qztd}. A total of 315 dust clouds are identified, with two sightlines having two clouds each as described in Section~\ref{sec:level}, and their central positions and boundaries are determined. This work primarily focuses on relatively diffuse high Galactic latitude dust clouds, including many dust clouds with $|b|>40^{\circ}$. Table~\ref{table1} provides the relevant information of the clouds.

\begin{figure*}
\centering
\includegraphics[width=1\textwidth,angle=0]{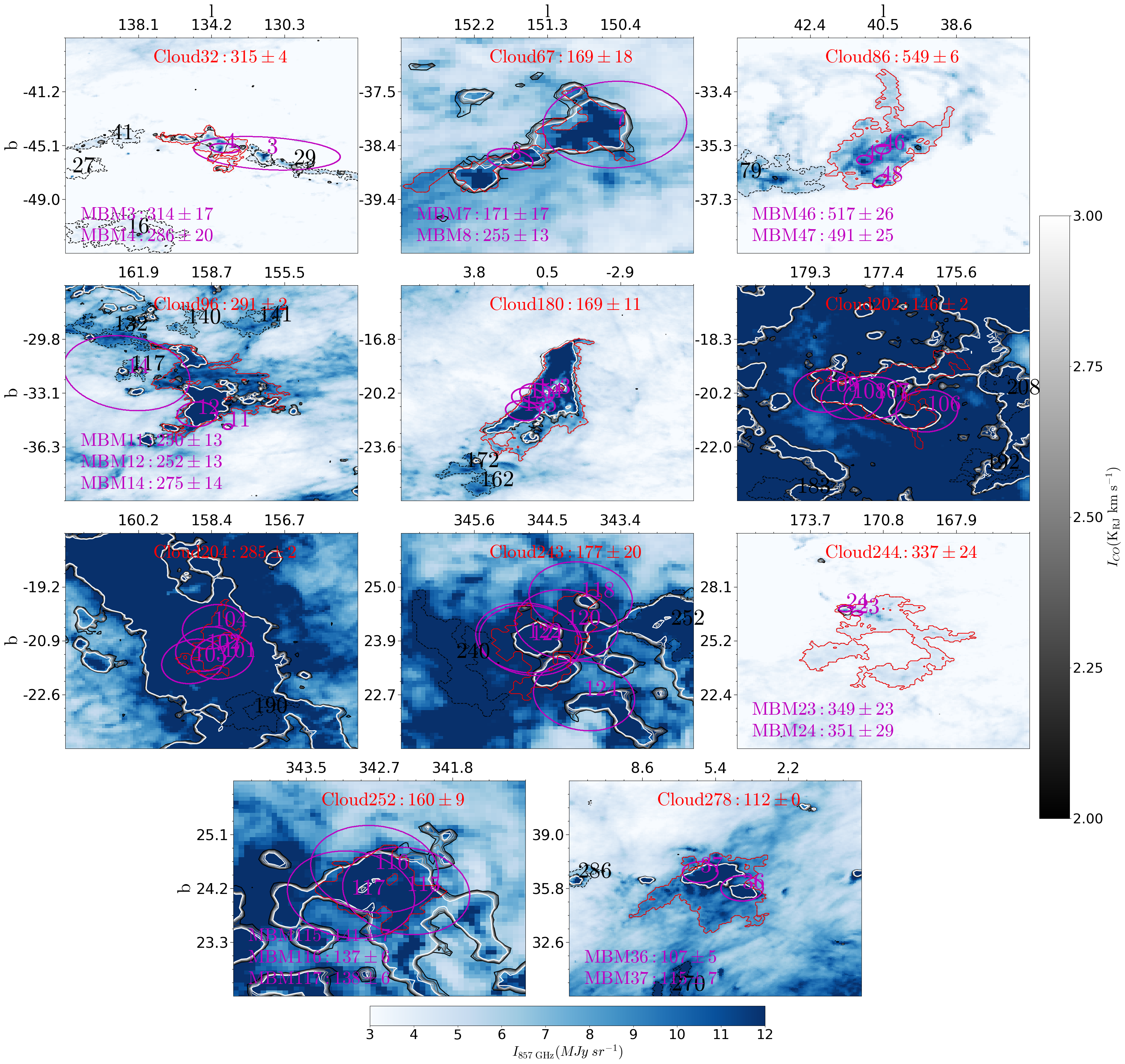}
\caption{The case of one cloud from hierarchical clustering corresponding to multiple MBM molecular clouds. The red solid lines and cloud numbers represent the dust cloud, while the magenta ellipses and numbers indicate the MBM clouds. The dashed black lines and numbers show other dust clouds identified in this study. The distances and errors for this work are shown in the top corner, and the distances and errors for the MBM clouds \citep{2019ApJ...879..125Z} are in the bottom left. The grayscale contour plot shows the CO data from the Planck survey.}
\label{cmbj}
\end{figure*}

To compare with the MBM molecular clouds, which are also in the high-latitude area, this work matches the clouds based on their their central coordinates and radius. The comparison result is listed in Table~\ref{table2}. There are 11 clouds identified in this study that cover multiple MBM molecular clouds fully or partly, and Figure~\ref{cmbj} shows the regions of these 11 clouds alongside the corresponding MBM molecular clouds. The MBM molecular clouds are commonly located in the densest regions of dust clouds. It is evident that the main bodies of the MBM clouds fall within the identified dust cloud boundaries of this study for Cloud 86, 244, 252, and 278 the difference of distance with the MBM clouds is either within or close to the margin of error. In these cases, the MBM clouds represent the densest areas within the dust cloud. However, there is some discrepancy in the distance between the MBM clouds and Cloud 32 and 96 in this study. By checking the neighbouring clouds, it is found that the main part of MBM3 corresponds to Cloud 29 instead of Cloud 32, and MBM14 corresponds to Cloud 117. In addition, in the Cloud 67 region, there is a significant difference between the distance to MBM 7 and MBM 8. On the other hand, apart from the situation where one dust cloud corresponds to multiple MBM clouds, there are also cases where one MBM cloud overlaps with multiple dust clouds. As shown in Table~\ref{table2}, in addition to the previously mentioned MBM 3 and MBM 14, MBM 16，MBM53 and MBM56 also overlap with multiple dust clouds. Upon examination, it is found that the center of MBM 16 is in closer proximity to the dense region of Cloud 73, the center of MBM 53 is closer to the dense region of Cloud 104, while the center of MBM 56 is closer to the dense region of Cloud 155.

\section{Distance measurement}
To determine cloud distances, we select a research region larger than each cloud and perform a local Gaussian fit on the number distribution of different $I_{857 \rm GHz}$ values. We use the region with $I_{857 \rm GHz}$ between $\mu-\sigma$ and $\mu$ as a reference for the diffuse interstellar medium (ISM). We first fit the ISM to determine its extinction variation with distance. Then, we apply an extinction-jump fit to the cloud region to obtain the cloud's distance and extinction. This method avoids premature jumps caused by insufficient foreground sources \citep{2021ApJS..256...46S}. To better reference cloud distances, we classify the results into four levels based on their jump characteristics and compare the distances of the most reliable clouds with results from other studies.

\subsection{Selection of reference region}

\begin{figure*}
\centering
\includegraphics[width=1\textwidth,angle=0]{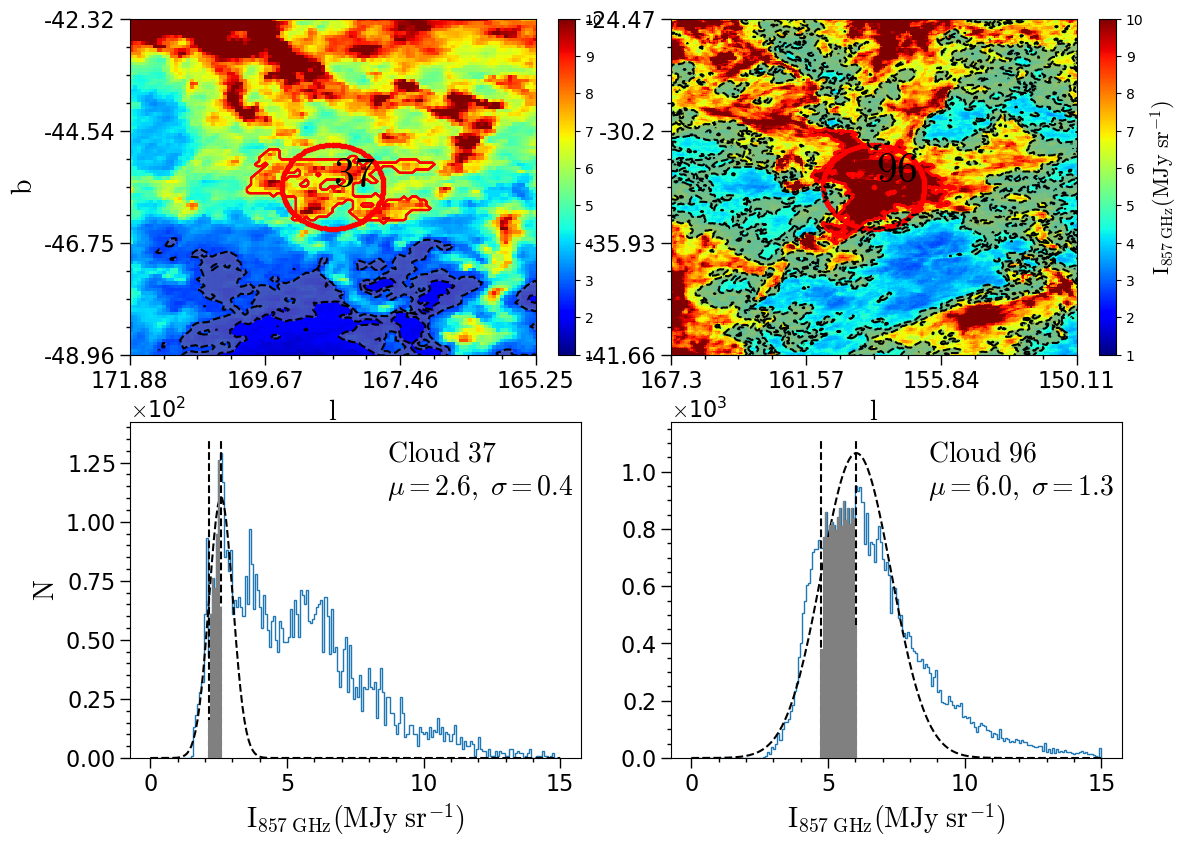}
\caption{The study region (top) and the corresponding histogram (bottom) of $I_{857}$ GHz ($\rm MJy sr^{-1}$) in cloud 37 (left) and cloud 96 (right). Within the study region, the gray filled region bounded by a black dashed line represents the reference region, while the green-yellow-red region enclosed by the red solid line corresponds the cloud region. The red solid circle is a circular shape of equal angular size to the cloud region. In the histogram, the blue steps represent the distribution of sources in the study region. the black dashed curve represents the local Gaussian fit, the gray bars represent sources in the reference region.}
\label{region}
\end{figure*}

A reference region should be selected to represent the variation of extinction when no cloud is present. For this purpose, a study region is selected for each cloud. Initially, a square area is selected with the center of the cloud. The side length of the square is set to be four times the cloud's equivalent angular diameter. However, when the observed intensity ($I_{857 \rm GHz}$) does not show a significant decrease within this region, the side length is progressively expanded by a factor of $m$ times the cloud's angular diameter until an appropriate reference region can be identified. The procedure of selecting the reference region is illustrated by two examples, Cloud 37 and Cloud 96, in Figure~\ref{region}.

The distribution of $I_{857 \rm GHz}$ is modeled by a Gaussian function to estimate the median ($\mu$) and standard deviation ($\sigma$) in the study region. Subsequently, the region with $I_{857 \rm GHz}$ between $\mu-\sigma$ to $\mu$ is designated as the reference region. This region is represented by the black dashed line and gray histogram in the upper and lower panels of Figure~\ref{region}, respectively. Simultaneously, the region identified by the Dendrogram `leaves' (represented by the red dashed line and histogram in the upper panel of Figure~\ref{region}) is selected as the cloud region.

\subsection{Distance determination}
The color excess is measured toward individual stars, some in front of and some behind the cloud, allowing us to trace the cloud's distance and extinction through the extinction jump. To accurately determine the distance and extinction of these dust clouds, we employ the extinction-jump model which has been applied to numerous supernova remnants and molecular clouds for distance and extinction measurements \citep{2017MNRAS.472.3924C, 2018ApJ...855...12Z, 2019MNRAS.488.3129Y, 2020ApJ...891..137Z, 2020MNRAS.493..351C,2021ApJS..256...46S,2022MNRAS.511.2302G}.
In brief, the total color excess along the line of sight towards the cloud can be described as the sum of two components: the dominant color excess of the cloud $E^{\rm MC}(d)$, and the color excess caused by the diffuse interstellar medium $E^{\rm DISM}(d)$. In this work, the distance and extinction of the dust clouds are determined by basically following the methodology proposed by \citet{2021ApJS..256...46S}. For the diffuse interstellar medium in the reference region, the interstellar extinction is expected to exhibit a smooth increase with distance. Meanwhile, the cloud region presents a jump at the distance of the cloud due to the much higher dust density of the cloud. The dominant color excess of the cloud $E^{\rm MC}$(d) is described by,
\begin{equation}\label{JUMP1}%
E^{\rm MC}(d)=\Delta E^{\rm MC} \times [1+erf( \frac{d-d_c}{\sqrt{2}\delta d})]
\end{equation}
where $\Delta E^{\rm MC}$ represents the jump magnitude of the color excess of the cloud.  Specifically it can be $\Delta E^{\rm MC}_{\rm G{BP},G_{RP}}$ between the Gaia BP and RP band, or $\Delta E^{\rm MC}_{\rm NUV,G{BP}}$ between the GALEX/NUV and Gaia/BP band. $d_c$ and $\delta d$ are the distance to the center and the thickness of the cloud, respectively.

The color excess of the diffuse interstellar medium $E^{\rm DISM}(d)$ is described by,
\begin{equation}\label{JUMP2}%
E^{\rm DISM}(d)=E^{\rm 0} \times(1-\exp {\frac{-d}{h'}})
\end{equation}
where the dust density decreases exponentially with the vertical distance from the Galactic plane along the cloud sightline at high Galactic latitude. The parameter $E^{\rm 0}$ represents the cumulative color excess, and $h' \times \sin(b)$, where $b$ is the latitude of the cloud, denotes the scale height ($h$) of the dust disk along the line of sight.

\begin{figure*}
\centering
\includegraphics[width=1\textwidth,angle=0]{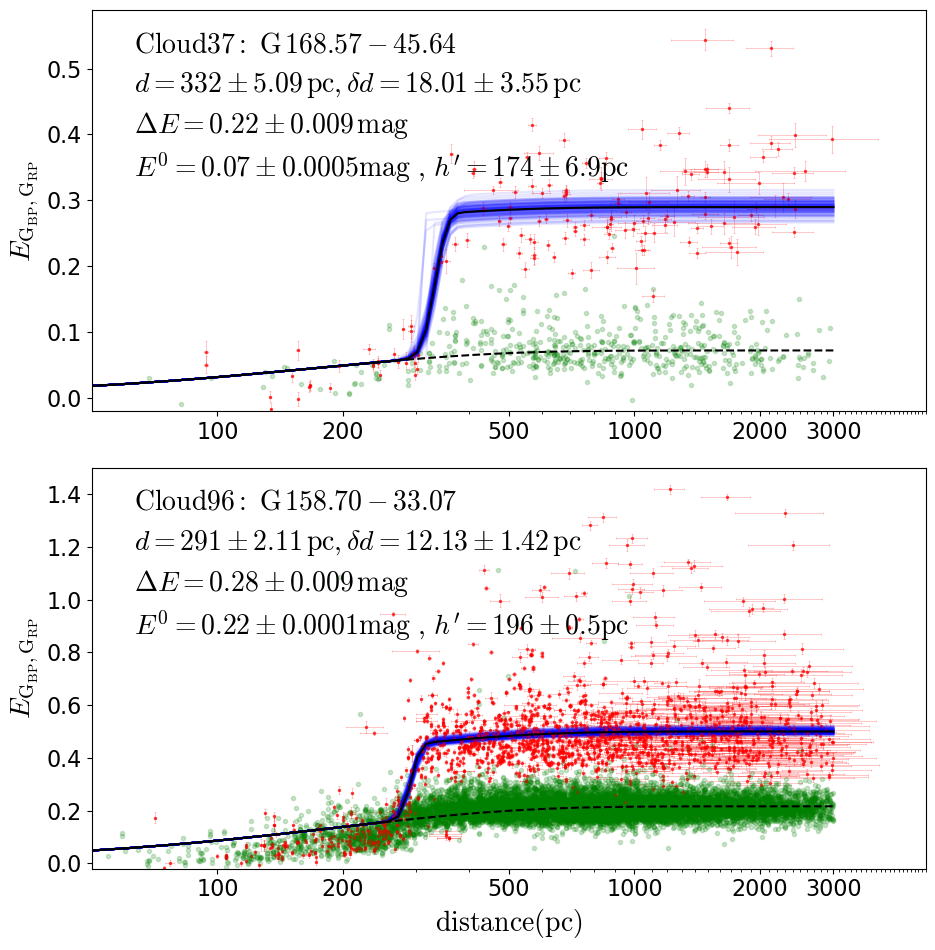}
\caption{The fitting of the color excess, $E_{\rm G_{BP},G_{RP}}$, variation with distance for stars in the reference (green dots) and cloud (red dots) regions is performed using the extinction-distance model (Eqs. \ref{JUMP1}, and \ref{JUMP2}). The solid black line represents the best fit for Eqs. \ref{JUMP1}, while the dashed black line represents the best fit for Eqs. \ref{JUMP2}. The blue lines represent the best fit for Eqs. \ref{JUMP1} among 100 randomly generated star samples. The derived parameters are displayed in the upper left corner, where  `d' and  `$\delta d$' represents the distance and thickness,  `$\Delta$E' corresponds to the color excess jump ($\Delta E^{\rm MC}_{\rm G{BP},G_{RP}}$), $E^{\rm 0}$ denotes the foreground parameters of the dust cloud ($E_{\rm G_{BP},G_{RP}}^{\rm 0}$), and $h'$ represents $h'_{\rm G_{BP},G_{RP}}$.  (An extended version of this figure for all the studied clouds (242 images) is available in the online journal)}
\label{op}
\end{figure*}

\begin{figure*}
\centering
\includegraphics[width=1\textwidth,angle=0]{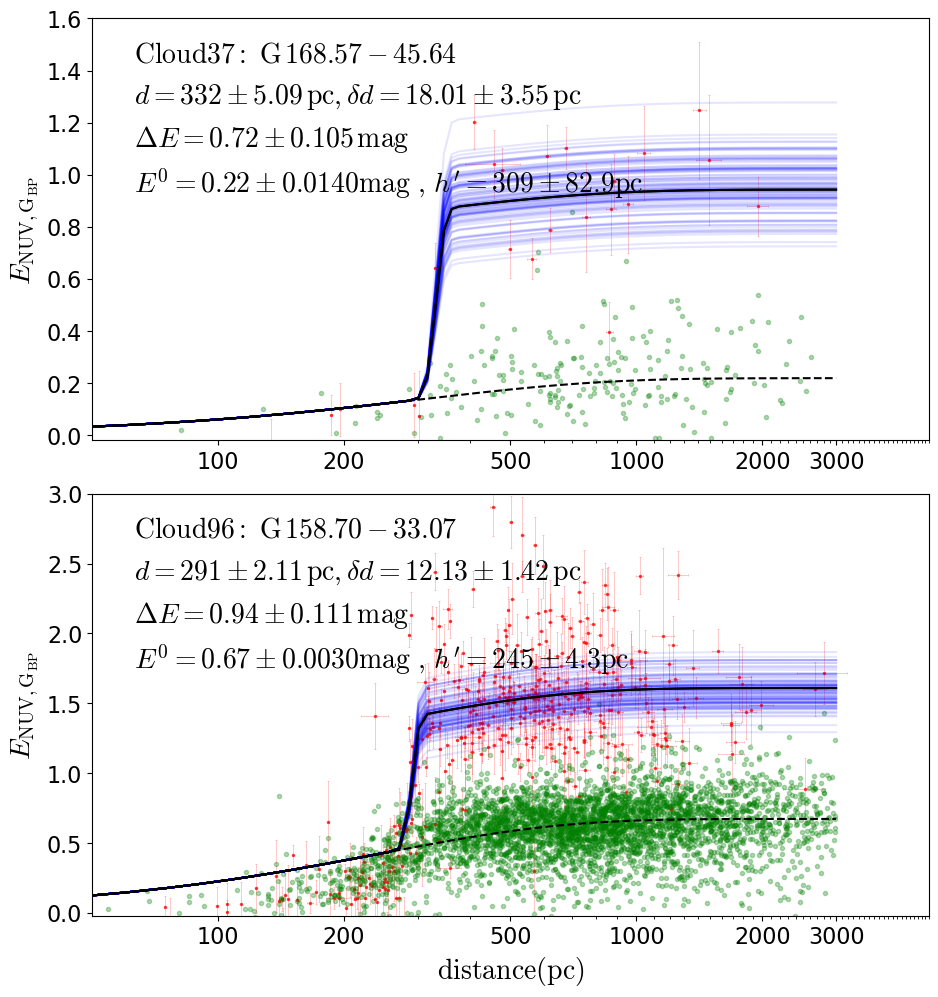}
\caption{The same as Figure~\ref{op}, but for $E_{\rm NUV,G_{BP}}$.}
\label{uv}
\end{figure*}

The model fitting is performed using the Markov Chain Monte Carlo (MCMC) procedure \citep{2013PASP..125..306F}. The stellar extinction variation with distance is fitted separately for the reference region and the cloud region. For the optical band, fitting the reference region can obtain the foreground parameters $E^{0}_{\rm G_{BP},G_{RP}}$ and $h'_{\rm G_{BP},G_{RP}}$, and the cloud region fitting can obtain the cloud distance $d$, the extinction jump $\Delta E^{\rm MC}_{\rm G_{BP},G_{RP}}$, and the cloud thickness $\delta d$. For the UV band, fitting the reference region can obtain the foreground parameters $E^{0}_{\rm NUV,G_{BP}}$ and $h'_{\rm NUV,G_{BP}}$, and the cloud region for the extinction jump $\Delta E^{\rm MC}_{\rm NUV, G_{BP}}$. To avoid the MCMC algorithm getting stuck in a local minimum, we not only consider the errors estimated by MCMC but also generate 100 sets of samples for each cloud based on the errors in source distance and extinction. The fitting algorithm is then applied to all the samples, and the errors are estimated using the root mean square (RMS) of the standard deviation of the best-fit parameters and the errors provided by MCMC. The model fitting for Cloud 37 and Cloud 96, with reference and cloud sources, is presented as an example in Figure~\ref{op} and Figure~\ref{uv}. The green dots in the figures represent the sources in the reference region used to determine the variation of extinction with distance in the diffuse medium, the red dots represent the sources in the cloud region used to determine the distance and extinction of the cloud, and the blue lines represent the best-fit lines for 100 sets of samples. The key parameters derived from the modeling, along with their uncertainties, are displayed in the upper left corner of the figures. 

\subsection{Distance accuracy}\label{sec:level}

\begin{figure*}
\centering
\includegraphics[width=1\textwidth,angle=0]{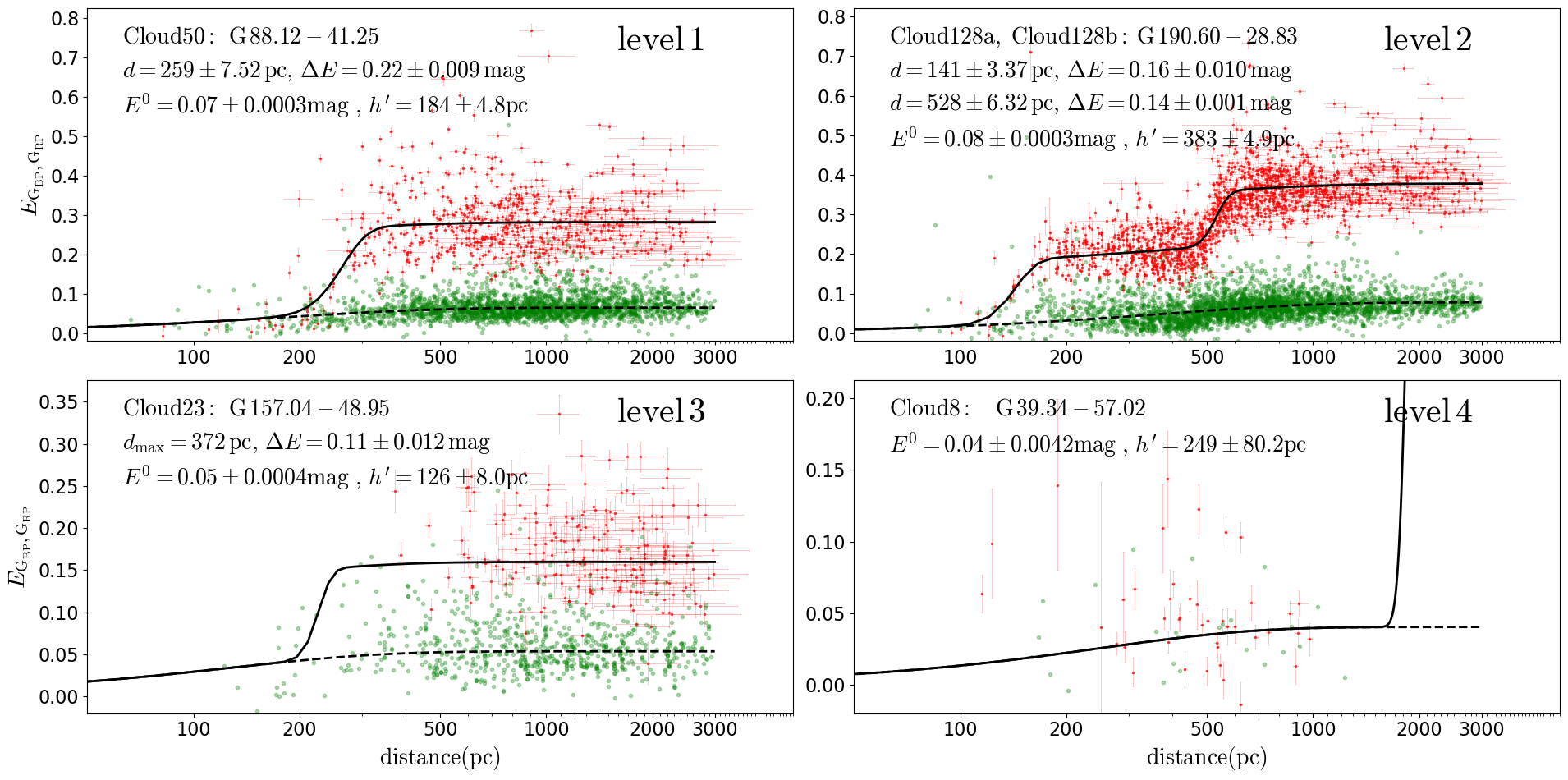}
\caption{The optical extinction-jump diagram for 4 level cloud. level 1: clouds with  exemplary jump; level 2: clouds with two jumps; level 3: clouds with significant uncertainty and lack of foreground sources; level 4: clouds with no identifiable jump.}
\label{op1}
\end{figure*}

\begin{figure*}
\centering
\includegraphics[width=1\textwidth,angle=0]{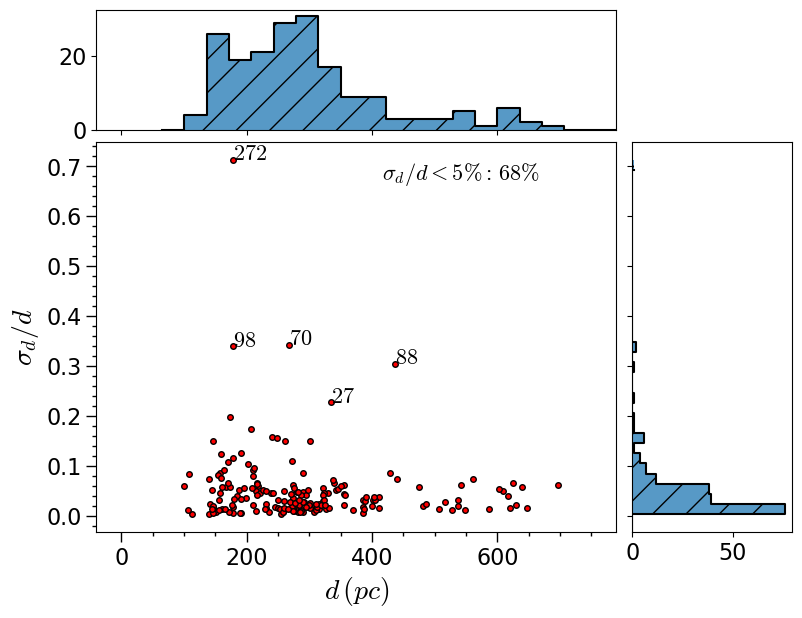}
\caption{The distribution of the distances (d) and their relative uncertainties ($\sigma_d/d$). The proportion of clouds with $\sigma_d/d<5\%$ are presented in the upper right corner of the panel.}
\label{distr}
\end{figure*}

Upon a preliminary visual examination, it is evident that the fitting demonstrates a high degree of concordance with the measurements across a majority of cases. To facilitate a more precise discourse on the cloud distance, the clouds are divided into four levels according to the jump features, as depicted in Figure~\ref{op1}. The Level 1 has exemplary jumps, wherein an ample number of foreground and cloud sources provides discernible leaps. The Level 2 refers to the presence of two discontinuities in the sightline, indicating the existence of two clouds along that sightline. The two clouds in the same sightline are labelled by "a" (closer) and "b" (further), respectively. The Level 3 denotes the jumps of an uncertain nature, wherein an almost negligible number of foreground stars precludes a definitive determination of the precise location of the cloud jump, and thus only an upper limit on the distance can be provided. The Level 4 corresponds to the absence of clear jumps, where the extinction does not exhibit a noticeable jump with distance. Consequently, Level 1, 2, 3 and 4 encompasses 188, 2, 22 and 30 objects respectively. Table~\ref{table1} presents the the relevant parameters for these clouds (An extended version of Figure~\ref{op} (242 images) and the complete Table~\ref{table1} for all the sample clouds are available in the online journal.) The distances of Level 1 and 2 clouds are generally reliable, and their distribution with the uncertainties for Level 1 and 2 clouds is displayed in Figure~\ref{distr}. It can be observed that the distance range of the clouds is primarily between 150-400 pc, and approximately 79$\%$ of the cloud distance error is less than 5$\%$. The later analysis uses only the data of the Level 1 and 2 clouds. 

\subsection{Comparison of distance}\label{level}

\begin{figure*}
\centering
\includegraphics[width=1\textwidth,angle=0]{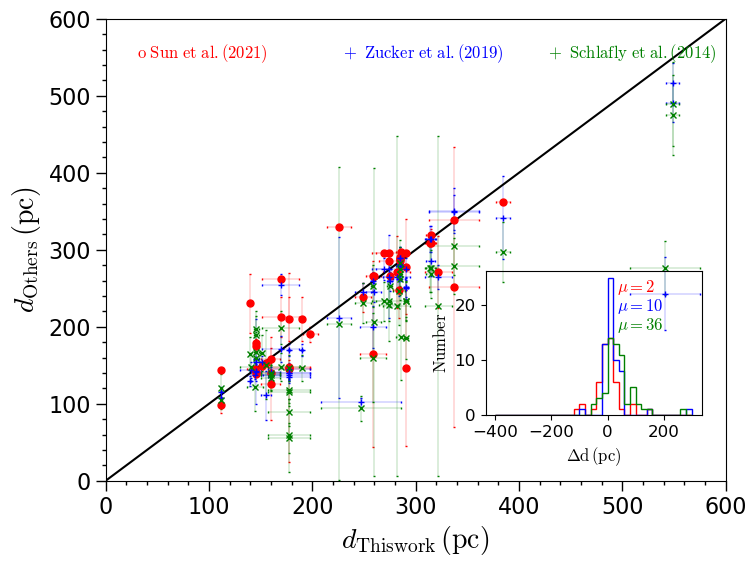}
\caption{The comparison of the distances of dust clouds with those derived by \citet{2014ApJ...786...29S}, \citet{2019ApJ...879..125Z}, and \citet{2021ApJS..256...46S}. The inset shows the distribution of the differences between our results and theirs.}
\label{comp}
\end{figure*}

There are 53 clouds used in our analysis examined by both \citet{2014ApJ...786...29S} and \citet{2019ApJ...879..125Z}, 59 clouds investigated by \citet{2021ApJS..256...46S}. The comparison of the distances in this work is presented in Figure~\ref{comp}. It can be seen that the distances to relatively far clouds are slightly larger compared to other works, but consistent with our previous results \citep{2021ApJS..256...46S}. his work yields systematically larger distances than \citet{2014ApJ...786...29S}, primarily due to accounting for molecular cloud thickness and separately considering the rise in foreground color excess with distance, which prevents premature jumps in some clouds \citep{2021ApJS..256...46S}. Overall, the results show good consistency with previous works, and there is a better agreement among this work, \citet{2019ApJ...879..125Z} and \citet{2021ApJS..256...46S}.

\section{Discussion}

\subsection{Tracing the Local Bubble}

The Local Bubble (LB) is a vast cavity within the Milky Way galaxy, located near the Sun. It extends from about $100-300$ pc away. The LB is characterized by low gas density, elevated temperature, and a higher abundance of ionized gas. It is believed to have formed as a result of supernova explosions \citep{2009ApJ...697L.158D,2022Natur.601..334Z}. \citet{2020A&A...636A..17P} use three-dimensional extinction map to extract the structure of the LB shell in the local interstellar medium and successfully fit a magnetic field model to observational data. It presents a novel approach for modeling the local Galactic magnetic field and dust polarized emission, providing valuable insights for studying polarized foregrounds in cosmic microwave background research. \citet{2022Natur.601..334Z} use new data from the Gaia space mission to analyze the 3D positions, shapes, and motions of dense gas and young stars near the Sun, revealing that star-forming complexes lie on the surface of the Local Bubble. The findings suggest that the Local Bubble originated from a burst of stellar birth and subsequent supernova explosions about 14 million years ago, leading to the expansion of the bubble and the formation of nearby clouds, supporting the theory of supernova-driven star formation. 

\begin{figure*}
\centering
\includegraphics[width=1\textwidth,angle=0]{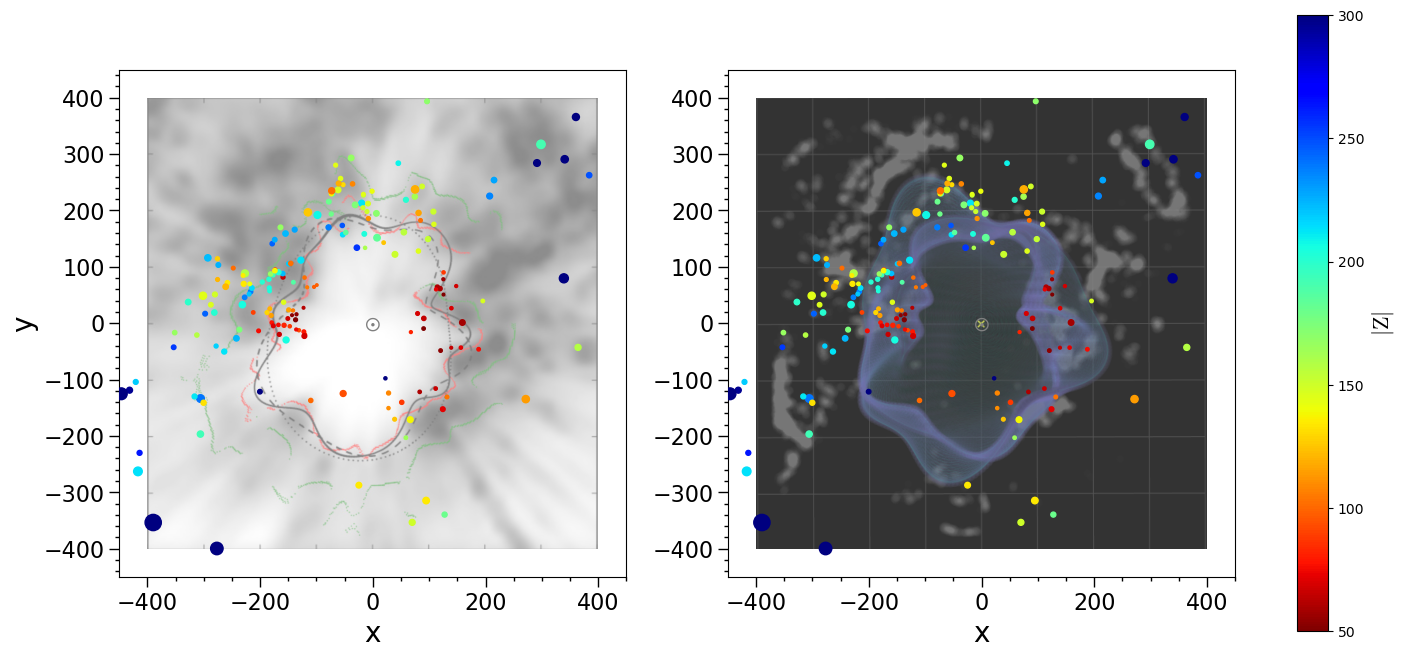}
\caption{The distribution of dust clouds is shown in a Cartesian coordinate system, with the colorbar indicating the Z-coordinate. The background image displays the contour of the \citet{2020A&A...636A..17P} bubble (left panel) and the \citet{2022Natur.601..334Z} bubble (right panel).}
\label{bub}
\end{figure*}

The high latitude dust clouds in this work are very nearby and can serve as a good tracer for the LB. The distribution in the Galactic XY plane of the clouds whose distances are well determined are displayed in Figure~\ref{bub}, where the color bar codes for the vertical distance $Z$.  It can be seen that there are indeed very few dust clouds inside the sphere of radius equal to 100\,pc. The comparison with the LB boundaries of \citet{2020A&A...636A..17P} (left panel of Figure~\ref{bub}) and \citet{2022Natur.601..334Z} (right panel of Figure~\ref{bub}) indicates that these clouds also outline the boundaries of the LB, in agreement with the two previous works with a better consistency with that of \citet{2022Natur.601..334Z}.

\subsection{Physical parameters of the clouds}

By integrating  all pixels belonging to the cloud, the solid angle ($\Omega$) subtended by the cloud is calculated as follows:

\begin{equation}
\Omega = \sum^N_{i=0}\Delta l \Delta b\ {\rm cos}\ b_i,
\end{equation}
Where $i$ is the pixel index, $\Delta l = \Delta b = 4'$ is the pixel angular width, and $b_i$ is the Galactic latitude of the $i$th pixel.

Given the distance ($d_0$) to the cloud, the area ($S$) and the equivalent linear radius ($r$) of the cloud can be expressed in units of pc$^{-2}$ and pc, respectively, as follows:

\begin{equation}
S= \Omega d_0^2 ,
\end{equation}

and

\begin{equation}
r= \sqrt{\frac{S}{\pi}} ,
\end{equation}

This work estimates the gas cloud mass from neutral hydrogen (HI) and molecular hydrogen ($\rm H_2$) observations in the area identified for the dust cloud using the method similar to \citet{2018ChA&A..42..213L}. The HI column density data is taken from the HI4PI survey \citep{2016A&A...594A.116H} and estimated by integrating spectroscopic data in velocity. Since molecular hydrogen lacks easily detectable spectral lines, we use carbon monoxide (CO) molecules as a tracer for the $\rm H_2$ column density, as is commonly done. The integrated intensity of the CO (J=1$\rightarrow$0) line is taken from the Planck survey \citep{2014A&A...571A..13P}. We adopt a constant CO integrated intensity to $\rm H_2$ column density conversion factor of $X_{\text{CO}} = 2.0 \times 10^{20} \text{cm}^{-2} \cdot \left[\text{K km s}^{-1}\right]^{-1}$ \citep{2013ARA&A..51..207B}. The column density of $\rm H_2$ can be represented as $N_{\rm H_2} = X_{\rm CO} \times W({\rm CO})$, where $W({\rm CO})$ is the integrated intensity of CO in units of K km/s. The gas mass of the cloud can be represented as，
\begin{equation}
M_{gas}= 1.36\sum^N_{i=0}(N_{HI}+2N_{\rm H_2})m_{\rm H} \Delta l \Delta b\ {\rm cos}\ b_i,
\end{equation}
where a multiplication factor of 1.36 is used to account for helium, $N_{\rm HI}$ and $N_{\rm H_2}$ are the column densities of HI and $\rm H_2$，$m_{\rm H} = 1.67 \times 10^{-24}$ g is the mass of a hydrogen atom.

The interstellar extinction curve of the Milky Way suggests that the interstellar dust is not composed of a single component, and its size distribution also has a certain range. As a result, dust models have been constructed to constrain the size and composition of interstellar dust. According to the WD01 model \citep{2001ApJ...548..296W}, the dust mass absorption coefficient is 2.8 $\times$ 10$^{4}$ mag $\cdot$ cm$^{2}$ $\cdot$ g$^{-1}$ for the $V$ band. Based on this, the dust mass column density per mag for ${A_{\rm V}}$ is $m_{dust}=3.57 \times 10^{-5}$ g $\cdot$ cm$^{-2}$ $\cdot$ mag$^{-1}$ \citep{2013ApJ...770...27N}. This work further utilizes the WD01 model to derive the dust mass directly from the optical extinction $A_{\rm V}$,
\begin{equation}
M_{dust}= A_{\rm V} \times m_{dust} \times S,
\end{equation}
Where  $A_{\rm V}$ is the optical extinction of the cloud, which is calculated as $2.2E_{\rm G{BP},G_{RP}}$ using the extinction law of \citet{2021ApJS..256...46S} and $m_{dust}$ is $3.57 \times 10^{-5}$ g $\cdot$ cm$^{-2}$ $\cdot$ mag$^{-1}$ as mentioned above. 

For $M_{dust} \ll M_{gas}$, the surface mass density $\Sigma$ of the molecular cloud is then calculated as,
\begin{equation}
\Sigma= M_{gas}/S,
\end{equation}
The gas-to-dust mass ratio (GDR) is,
\begin{equation}
{\rm GDR}= M_{gas}/M_{dust},
\end{equation}

\begin{figure*}
\centering
\includegraphics[width=1\textwidth,angle=0]{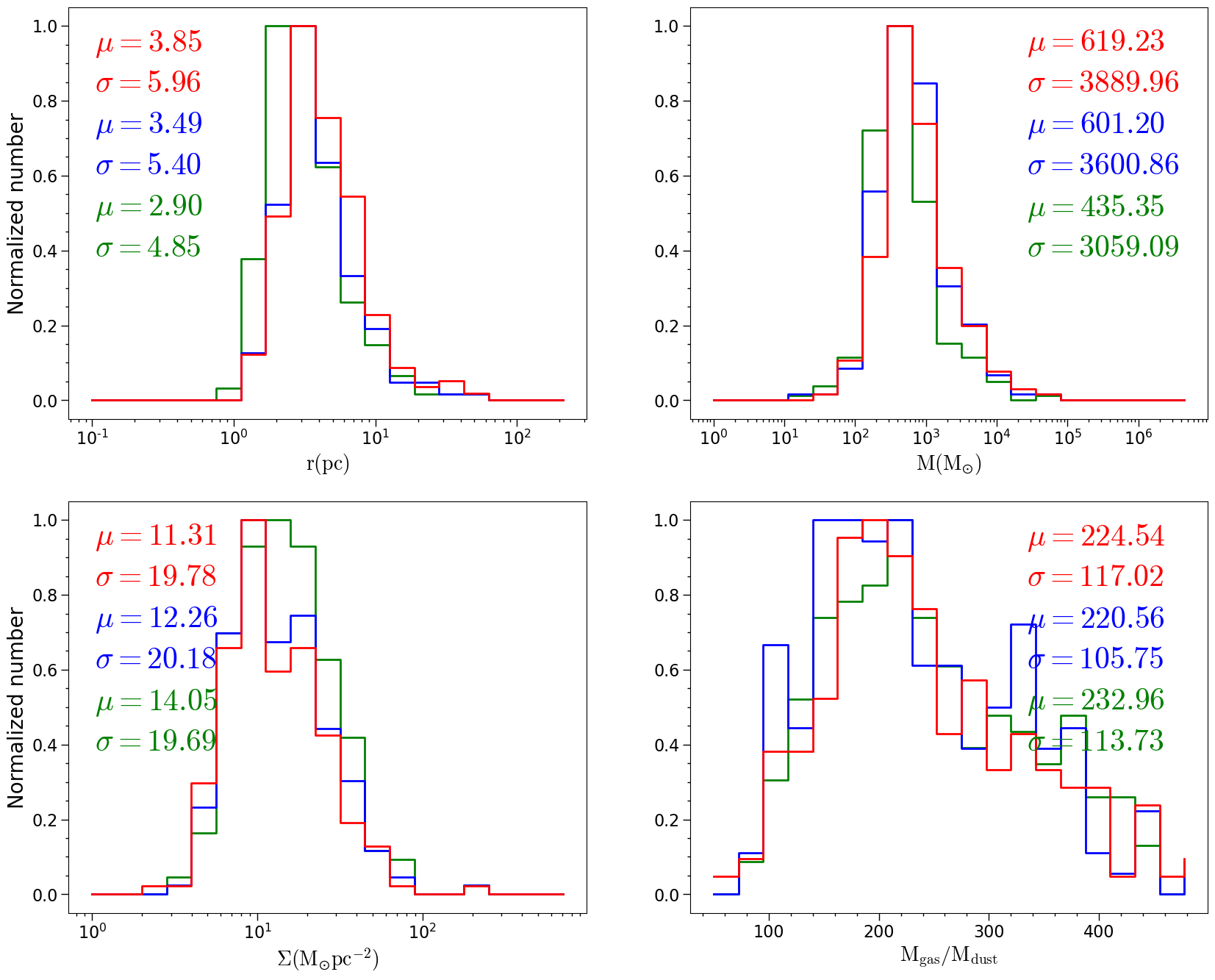}
\caption{Normalized histograms distributions of the physical properties, including line radius (upper-left panel), masses (upper-right panel), surface mass densities (bottom-left panel), GDR (bottom-right panel) of the well-defined distance clouds. The red, blue, and green lines represent the parameter distributions for min\_npix=225 (this work), min\_npix=144, and min\_npix=81, respectively. The median and standard deviation are presented in the top corner of every panel in matching colors.}
\label{mgx}
\end{figure*}

Figure~\ref{mgx} presents the histograms of these parameters of the clouds. The cloud radius primarily concentrates between 2 and 15 pc, with a median value of approximately 4 pc. The mass primarily concentrates between 150 and 6000 $\rm M_\odot$, with a median value of approximately 620 $\rm M_\odot$. The surface mass density primarily concentrates between 5 and 40 $\rm M_\odot pc^{-2}$, with a median value of approximately 11 $\rm M_\odot pc^{-2}$. The GDR primarily concentrates between 100 and 450, with a median value of approximately 225 for clouds with $\Delta E^{\rm MC}_{\rm G{BP},G_{RP}}>0.05 mag$. The GDR can vary depending on the interstellar environment and dust composition \citep{1978ApJ...225...40B,2013ApJ...777....5S}. The derived GDR in this work is slightly higher than the widely accepted average value range of 100-150 for the Milky Way \citep{1978ppim.book.....S,1983QJRAS..24..267H,1984ApJ...285...89D}, which is a value range for the diffuse interstellar medium. However, observed GDR values exhibit significant scatter, with \citet{1978ppim.book.....S} reporting a broad range from 20 to 700. Using the same dust model, the GDR derived in this work aligns well with \citet{2018ChA&A..42..213L}, who reported an average GDR of 260 for the Orion, Taurus, and Polaris star-forming regions. The results from other galaxies observations have also shown large variations in GDR. \citet{1990ApJ...359...42D} measured the inner disk gas mass to warm dust mass ratio in spiral galaxies at 1080 $\pm$ 70, and \citet{1991ARA&A..29..581Y} derived an average GDR of 600 for molecular clouds using IRAS data, both significantly higher than the typical value for the Galaxy. \citet{2014MNRAS.444L..90B} analyzed NGC 5485, an early-type galaxy with significant cold dust, and found a GDR ratio upper limit of 14.5, much lower than the Milky Way's. 

In this work, the minimum angular size of catalogued clouds is 1 deg$^2$ (min\_npix = 225), which constrains the distribution of cloud radius, masses, surface mass densities, and other parameters. To understand the impact of min\_npix on parameter distributions, we also simulated distributions for min\_npix = 144 (0.64 deg$^2$) and min\_npix = 81 (0.32 deg$^2$) in Figure~\ref{mgx}. The variation in min\_npix can affect the parameter distribution to some extent, but the impact is not significant. As min\_npix decreases, the lower limits and medians of the radius and mass distributions show a slight decrease, while the surface mass density distributions experience a slight increase, which is because smaller clouds and denser regions of partial clouds are identified as clouds. Meanwhile, the median and dispersion of the GDR distribution show minimal change.

\subsection{The extinction and dust property of clouds}
\begin{figure}
\centering
\centerline{\includegraphics[scale=0.9]{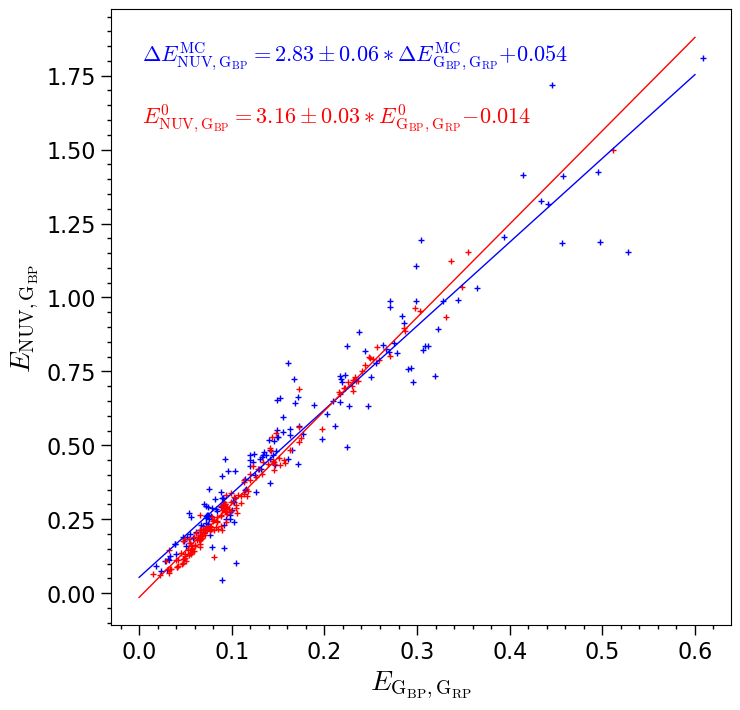}}
\caption{The relationship between $\Delta E^{\rm MC}_{\rm G{BP},G{RP}}$ and $\Delta E^{\rm MC}_{\rm NUV,G{BP}}$ (blue plus sign and blue line), as well as $E^{\rm 0}_{\rm G{BP},G_{RP}}$ and $E^{\rm 0}_{\rm NUV,G{BP}}$ (red plus sign and red line),
\label{tbnh}}
\end{figure}

The optical color excess $\Delta E^{\rm MC}_{\rm G{BP},G_{RP}}$ is determined for 190 clouds, while the UV-optical color excess $\Delta E^{\rm MC}_{\rm NUV,G{BP}}$ is determined for 165 clouds. The relation between $E^{\rm MC}_{\rm G{BP},G_{RP}}$ and  $\Delta E^{\rm MC}_{\rm NUV,G{BP}}$ for the clouds is compared with $E_{\rm G_{BP},G_{RP}}^{\rm 0}$ and $E_{\rm NUV,G_{BP}}^{\rm 0}$ for the reference region in Figure~\ref{tbnh}. There is a tight linear relationship between the optical and UV color excess $E^{\rm 0}_{\rm NUV,G_{BP}}/E^{\rm 0}_{\rm G_{BP},G_{RP}}$, with a slope of 2.83 $\pm$ 0.06 and 3.16 $\pm$ 0.03 for the cloud and reference regions, respectively. The ratio agrees very well with the all-sky color excess ratio of 3.25 reported by \citet{2021ApJS..254...38S}.

\begin{figure*}
\centering
\includegraphics[width=0.8\textwidth,angle=0]{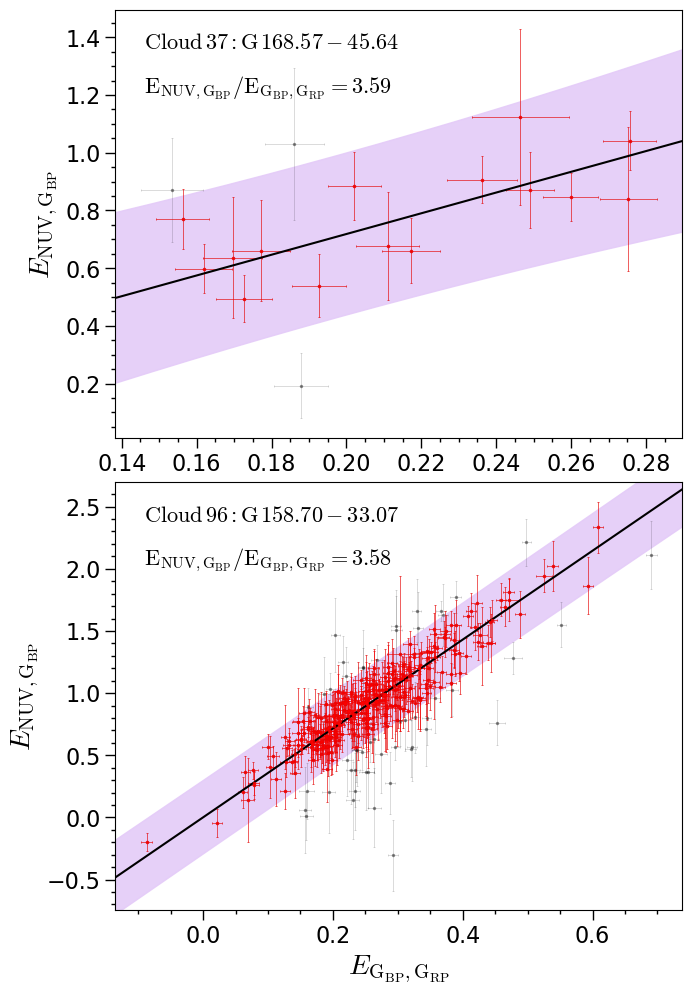}
\caption{The linear fitting for the color excess $E_{\rm NUV,G_{BP}}$ to $E_{\rm G_{BP},G_{RP}}$ in cloud 37 and cloud 96. Red and black dots represent sources within and outside the 95\% confidence interval, respectively. The black line represents the fitting curve, and the purple shadow represents the 95\% confidence interval.}
\label{ratio}
\end{figure*}

Parameters $\Delta E^{\rm MC}$ are derived by fitting Equation~\ref{JUMP1} to all stars toward a cloud, with the extinction jump magnitude corresponding to $\Delta E^{\rm MC}$. Due to selection effects, UV samples typically observe areas with lower cloud extinction than optical samples, even for the same cloud, often capturing more diffuse regions. Therefore, the ratio $\Delta E^{\rm MC}_{\rm NUV,G{BP}}/\Delta E^{\rm MC}_{\rm G{BP},G_{RP}}$ is not suitable for studying cloud dust properties, leading to an underestimated color excess ratio. Instead, the sources behind the cloud are all taken into account to determine the color excess ratio $E_{\rm NUV,G_{BP}}$/$E_{\rm G_{BP},G_{RP}}$ and then the dust property of a dust cloud. By subtracting the color excess of the diffuse interstellar medium at equal distance, the corresponding color excess of the cloud is obtained. By requiring at least three selected sources (N$\geq$3), The color excess ratios of 165 dust clouds are calculated using linear fitting with iterative 3$\sigma$ clipping \citep{2021ApJS..254...38S}. The results of Cloud 12 and Cloud 45 are displayed in Figure~\ref{ratio}, where $E_{\rm NUV,G_{BP}}$ and $E_{\rm G_{BP},G_{RP}}$ have a well-defined linear relationship. 

\begin{figure*}
\centering
\includegraphics[width=1\textwidth,angle=0]{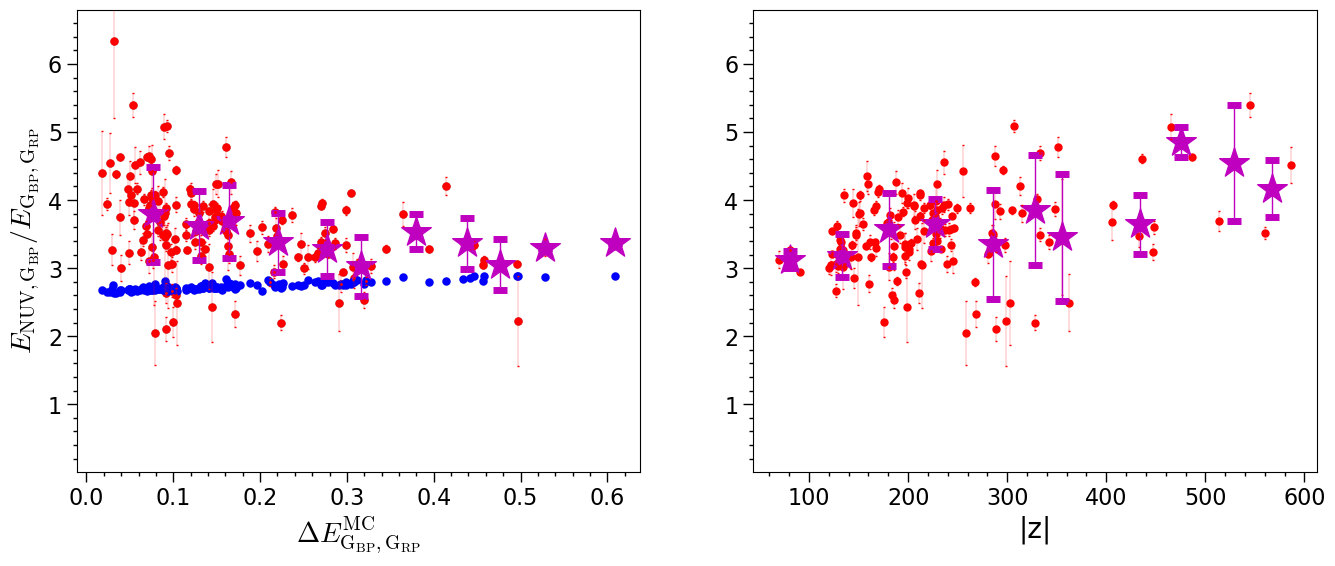}
\caption{Left panel: the change of color excess ratios $E_{\rm NUV,G_{BP}}$/$E_{\rm G_{BP},G_{RP}}$ with $\Delta E^{\rm MC}_{\rm G_{BP},G_{RP}}$. Right panel: the change of color excess ratios $E_{\rm NUV,G_{BP}}$/$E_{\rm G_{BP},G_{RP}}$ with Galactic plane distance $|z|$. Red dots are the results of the studied dust clouds and the purple asterisks are median color excess ratios of 0.05 $\Delta E^{\rm MC}_{\rm G_{BP},G_{RP}}$ interval (left panel) and 50 pc Galactic plane distance $|z|$ interval (right panel). The blue dots in the left panel are the simulation results considering the effective wavelength shift because of $\Teff$ and $\Delta E^{\rm MC}_{\rm G_{BP},G_{RP}}$. }
\label{prop}
\end{figure*}

We will utilize the Level 1 and Level 2 clouds to discuss the variation of color excess ratios ($E_{\rm NUV,G_{BP}}/E_{\rm G_{BP},G_{RP}}$) with respect to changes in optical extinction jump $\Delta E^{\rm MC}_{\rm G_{BP},G_{RP}}$ and Galactic plane distance $|z|$. The change of color excess ratios ($E_{\rm NUV,G_{BP}}/E_{\rm G_{BP},G_{RP}}$) with $\Delta E^{\rm MC}_{\rm G_{BP},G_{RP}}$  of these dust clouds are shown in the left panel of Figure~\ref{prop}. The blue points represent the theoretical color excess ratios for $R_{\rm V}$=3.1 \citep{1999PASP..111...63F} accounting for effective wavelength shifts. The theoretical color excess ratios increase with $\Delta E^{\rm MC}_{\rm G{BP},G_{RP}}$. The measured color excess ratios (red points and purple stars) are notably higher than the theory at lower $\Delta E^{\rm MC}_{\rm G{BP},G_{RP}}$. Higher UV color excess ratios correspond to smaller $R_{\rm V}$, indicating a larger fraction of small dust grains, consistent with less shielding in low extinction regions \citet{2021ApJS..256...46S}. Furthermore, the right panel of Figure~\ref{prop} illustrates the variation of the color excess ratio with respect to Galactic plane distance $|z|$. It can be observed that as the Galactic plane distance $|z|$ increases, the color excess ratio tends to increase overall. Clouds at larger Galactic distances tend to have lower extinction, and therefore contain a higher fraction of small dust grains. This is consistent with the observed variations in color excess ratios with  $\Delta E^{\rm MC}_{\rm G{BP},G_{RP}}$.

\section{Summary}

This work makes use of a clustering hierarchical algorithm to identify high-latitude dust clouds with the Planck 857 GHz dust emission map.  The distances to the identified clouds are precisely determined. In addition, the physical parameters and dust properties of these clouds are discussed.

The major results are following:

\begin{enumerate}

\item A total of 315 high-latitude dust clouds at $|b|>20\degr$ are identified using a clustering hierarchical algorithm. 

\item The distances of 190 dust clouds are precisely measured using the extinction-jump model, including 188 clouds with single jumps and 2 clouds with double jumps. For the common sources, the distances in this study agrees well with those of \citet{2019ApJ...879..125Z} and \cite{2021ApJS..256...46S}. The clouds with well-defined distance closely resemble the outline of the local bubble.

\item The basic physical parameters of the dust clouds, including the linear radius ($r$), mass ($M$), surface mass density ($\Sigma$), and gas-to-dust mass ratio (GDR) are calculated. The study of the color excess ratios of the dust clouds confirms that the dust clouds with larger Galactic distances and lower extinctions may have a greater proportion of small-sized dust particles.

\end{enumerate}

\section*{Acknowledgements}
We would like to thank the referee for providing us with detailed and constructive feedback that has significantly enhanced the quality of the manuscript. We thank Zhetai Cao, Jun Li, Ruoyi Zhang, and Xiaoxiao Ma for helpful discussions. This work is supported by NSFC through projects 12203016, 12173013, 12133002 and 11533002, Natural Science Foundation of Hebei Province No.~A2022205018, A2021205006, 226Z7604G, and Science Foundation of Hebei Normal University No.~L2022B33. W.Y.C. acknowledge the support form the science research grants from the China Manned Space Project. M.X.S. acknowledge the support of Physics Postdoctoral Research Station at Hebei Normal University. This work made use of the data taken by \emph{GALEX}, LAMOST, \emph{Gaia} and GALAH, Planck.

\facilities{\emph{GALEX}, LAMOST, \emph{Gaia}, GALAH, Planck}

\bibliographystyle{aasjournal}
\bibliography{spp}
\clearpage

\clearpage

\begin{deluxetable}{cccccccccccccccccc}
\rotate
\caption{\label{table1}The main information of the dust clouds}
\tablehead{\colhead{Cloud{\tablenotemark{a}}} & \colhead{$l${\tablenotemark{a}}} & \colhead{$b${\tablenotemark{a}}} & \colhead{$r${\tablenotemark{a}}} & \colhead{$d_{0}${\tablenotemark{a}}} & \colhead{$\delta d${\tablenotemark{a}}} & \colhead{$level${\tablenotemark{a}}} & \colhead{$E^0_{G_{BP},G_{RP}}${\tablenotemark{a}}} & \colhead{$h'_{G_{BP},G_{RP}}${\tablenotemark{a}}} & \colhead{$\Delta E^{\rm MC}_{G_{BP},G_{RP}}${\tablenotemark{a}}} & \colhead{m{\tablenotemark{a}}} & \colhead{$E^0_{NUV,G_{BP}}${\tablenotemark{a}}} & \colhead{$h'_{NUV,G_{BP}}${\tablenotemark{a}}} & \colhead{$\Delta E^{\rm MC}_{NUV,G_{BP}}${\tablenotemark{a}}} & \colhead{$M${\tablenotemark{a}}} & \colhead{$\Sigma${\tablenotemark{a}}} & \colhead{GDR{\tablenotemark{a}}} & \colhead{CERs{\tablenotemark{a}}}\\
\colhead{ } & \colhead{$(^{\circ})$}& \colhead{$(^{\circ})$}& \colhead{$pc$}& \colhead{$pc$} & \colhead{$pc$} & \colhead{ } & \colhead{$mag$} & \colhead{$pc$} & \colhead{$mag$} & \colhead{ } & \colhead{$mag$} & \colhead{$pc$} & \colhead{$mag$} & \colhead{$M_\odot$} & \colhead{$M_\odot pc^-2$}  & \colhead{ } & \colhead{ } \\
\colhead{(1)} & \colhead{(2)}& \colhead{(3)} & \colhead{(4)} & \colhead{(5)}  & \colhead{(6)} & \colhead{(7)}  & \colhead{(8)}  & \colhead{(9)}
& \colhead{(10)}  & \colhead{(11)}  & \colhead{(12)} & \colhead{(13)} & \colhead{(14)}
 & \colhead{(15)} & \colhead{(16)} & \colhead{(17)} & \colhead{(18)}
}
\startdata
${\rm Cloud}0$ & $97.42$ & $-87.97$ &  &  &  &  &  &  &  &  &  &  &  &  &  &  &  \\
${\rm Cloud}1$ & $125.52$ & $-71.14$ &  &  &  &  &  &  &  &  &  &  &  &  &  &  &  \\
${\rm Cloud}2$ & $134.29$ & $-69.03$ &  &  &  &  &  &  &  &  &  &  &  &  &  &  &  \\
${\rm Cloud}3$ & $127.48$ & $-69.90$ &  &  &  &  &  &  &  &  &  &  &  &  &  &  &  \\
${\rm Cloud}4$ & $115.44$ & $-68.84$ &  &  &  &  &  &  &  &  &  &  &  &  &  &  &  \\
${\rm Cloud}5$ & $139.19$ & $-66.27$ &  &  &  & $4$ & $0.05\pm0.0004$ & $64\pm11$ &  & $4$ &  &  &  &  &  &  &  \\
${\rm Cloud}6$ & $227.69$ & $-66.24$ &  &  &  &  &  &  &  &  &  &  &  &  &  &  &  \\
${\rm Cloud}7$ & $101.90$ & $-61.77$ & $5.8$ & $290\pm25$ & $14\pm12$ & $1$ & $0.04\pm0.0003$ & $154\pm7$ & $0.03\pm0.025$ & $4$ & $0.11\pm0.1606$ & $129\pm 47$ & $0.11\pm0.109$ & $265$ & $2.5$ & $323$ & $4.54$ \\
${\rm Cloud}8$ & $39.34$ & $-57.02$ &  &  &  & $4$ & $0.04\pm0.0042$ & $249\pm80$ &  & $4$ & $0.22\pm0.0535$ & $422\pm136$ & $1.53\pm1.029$ &  &  &  &  \\
${\rm Cloud}9$ & $87.15$ & $-50.43$ & $8.6$ & $238\pm11$ & $37\pm8$ & $1$ & $0.06\pm0.0002$ & $139\pm4$ & $0.08\pm0.010$ & $4$ & $0.16\pm0.0051$ & $170\pm 32$ & $0.29\pm0.093$ & $902$ & $3.9$ & $178$ & $4.07$ \\
${\rm Cloud}10$ & $107.31$ & $-53.76$ & $3.6$ & $307\pm3$ & $19\pm2$ & $1$ & $0.05\pm0.0005$ & $125\pm11$ & $0.24\pm0.008$ & $4$ & $0.11\pm0.0106$ & $121\pm 66$ & $0.88\pm0.143$ & $254$ & $6.2$ & $95$ & $3.78$ \\
${\rm Cloud}11$ & $190.92$ & $-52.78$ & $7.3$ & $259\pm13$ & $0\pm4$ & $1$ & $0.09\pm0.0004$ & $262\pm5$ & $0.04\pm0.010$ & $4$ & $0.22\pm0.0082$ & $243\pm 35$ & $0.13\pm0.178$ & $1009$ & $6.0$ & $544$ & $3.00$ \\
${\rm Cloud}12$ & $117.65$ & $-52.69$ &  &  &  &  &  &  &  &  &  &  &  &  &  &  &  \\
${\rm Cloud}13$ & $114.76$ & $-51.82$ & $5.1$ & $303\pm5$ & $21\pm6$ & $1$ & $0.05\pm0.0007$ & $137\pm20$ & $0.17\pm0.008$ & $4$ & $0.12\pm0.0184$ & $224\pm151$ & $0.73\pm0.156$ & $443$ & $5.5$ & $119$ & $4.26$ \\
${\rm Cloud}14$ & $95.43$ & $-52.02$ & $4.2$ & $259\pm2$ & $13\pm2$ & $1$ & $0.10\pm0.0003$ & $159\pm3$ & $0.15\pm0.009$ & $4$ & $0.31\pm0.0066$ & $201\pm 23$ & $0.65\pm0.153$ & $346$ & $6.3$ & $153$ & $4.24$ \\
${\rm Cloud}15$ & $108.75$ & $-52.12$ & $4.0$ & $270\pm4$ & $23\pm2$ & $1$ & $0.05\pm0.0007$ & $161\pm13$ & $0.30\pm0.010$ & $4$ & $0.13\pm0.0117$ & $124\pm 76$ & $0.99\pm0.103$ & $560$ & $11.1$ & $134$ & $3.85$ \\
${\rm Cloud}16$ & $138.74$ & $-51.44$ & $7.0$ & $272\pm5$ & $24\pm3$ & $1$ & $0.06\pm0.0002$ & $193\pm3$ & $0.09\pm0.008$ & $4$ & $0.15\pm0.0041$ & $173\pm 21$ & $0.34\pm0.095$ & $731$ & $4.7$ & $192$ & $4.12$ \\
${\rm Cloud}17$ & $72.13$ & $-49.60$ & $6.5$ & $198\pm7$ & $18\pm4$ & $1$ & $0.08\pm0.0002$ & $145\pm2$ & $0.07\pm0.009$ & $4$ & $0.25\pm0.0041$ & $169\pm 14$ & $0.23\pm0.118$ & $627$ & $4.7$ & $245$ & $3.63$ \\
${\rm Cloud}18$ & $205.48$ & $-50.30$ &  &  &  &  &  &  &  &  &  &  &  &  &  &  &  \\
${\rm Cloud}19$ & $151.11$ & $-50.60$ & $3.8$ & $287\pm13$ & $15\pm7$ & $1$ & $0.09\pm0.0006$ & $251\pm6$ & $0.05\pm0.008$ & $4$ & $0.29\pm0.0117$ & $273\pm 35$ & $0.18\pm0.164$ & $219$ & $4.8$ & $356$ & $3.97$ \\
${\rm Cloud}20$ & $144.74$ & $-50.26$ & $3.5$ & $286\pm5$ & $1\pm0$ & $1$ & $0.08\pm0.0007$ & $222\pm8$ & $0.13\pm0.008$ & $4$ & $0.24\pm0.0131$ & $199\pm 40$ & $0.46\pm0.095$ & $514$ & $13.5$ & $372$ & $3.57$ \\
${\rm Cloud}21$ & $95.82$ & $-50.05$ & $2.1$ & $209\pm17$ & $5\pm2$ & $1$ & $0.13\pm0.0007$ & $222\pm5$ & $0.14\pm0.010$ & $4$ & $0.42\pm0.0163$ & $225\pm 36$ & $0.37\pm0.241$ & $94$ & $7.1$ & $181$ & $3.02$ \\
${\rm Cloud}22$ & $152.71$ & $-49.70$ & $2.5$ & $245\pm4$ & $1\pm1$ & $1$ & $0.09\pm0.0007$ & $253\pm8$ & $0.05\pm0.011$ & $4$ & $0.29\pm0.0153$ & $303\pm 42$ & $0.16\pm0.088$ & $115$ & $5.7$ & $410$ & $4.35$ \\
${\rm Cloud}23$ & $157.04$ & $-48.95$ &  & $372$ &  & $3$ & $0.05\pm0.0004$ & $126\pm8$ & $0.11\pm0.012$ & $4$ & $0.14\pm0.0140$ & $389\pm127$ & $0.32\pm0.290$ &  &  &  &  \\
${\rm Cloud}24$ & $106.56$ & $-47.78$ & $3.6$ & $250\pm4$ & $1\pm5$ & $1$ & $0.05\pm0.0018$ & $236\pm23$ & $0.14\pm0.008$ & $4$ & $0.18\pm0.0208$ & $261\pm116$ & $0.48\pm0.103$ & $166$ & $4.1$ & $110$ & $3.91$ \\
${\rm Cloud}25$ & $165.50$ & $-46.79$ & $4.0$ & $327\pm4$ & $21\pm3$ & $1$ & $0.07\pm0.0004$ & $176\pm9$ & $0.15\pm0.009$ & $4$ & $0.20\pm0.0131$ & $252\pm 68$ & $0.48\pm0.124$ & $458$ & $9.0$ & $219$ & $3.71$ \\
${\rm Cloud}26$ & $151.57$ & $-46.54$ & $5.1$ & $277\pm7$ & $1\pm4$ & $1$ & $0.10\pm0.0005$ & $202\pm5$ & $0.15\pm0.011$ & $4$ & $0.29\pm0.0094$ & $215\pm 25$ & $0.53\pm0.198$ & $680$ & $8.3$ & $202$ & $3.75$ \\
${\rm Cloud}27$ & $141.65$ & $-47.06$ & $3.8$ & $334\pm76$ & $5\pm8$ & $1$ & $0.08\pm0.0008$ & $277\pm11$ & $0.02\pm0.015$ & $4$ & $0.21\pm0.0148$ & $221\pm 57$ & $0.09\pm0.144$ & $219$ & $4.9$ & $959$ & $4.40$ \\
${\rm Cloud}28$ & $162.67$ & $-46.71$ & $2.4$ & $280\pm3$ & $1\pm1$ & $1$ & $0.09\pm0.0009$ & $196\pm11$ & $0.16\pm0.009$ & $4$ & $0.29\pm0.0264$ & $366\pm102$ & $0.55\pm0.192$ & $192$ & $10.3$ & $227$ & $3.49$ \\
${\rm Cloud}29$ & $129.84$ & $-46.50$ & $4.8$ & $314\pm5$ & $1\pm1$ & $1$ & $0.07\pm0.0004$ & $201\pm5$ & $0.13\pm0.009$ & $4$ & $0.22\pm0.0068$ & $268\pm 33$ & $0.34\pm0.227$ & $874$ & $12.0$ & $341$ & $3.54$ \\
${\rm Cloud}30$ & $9.54$ & $-46.44$ &  &  &  & $4$ & $0.14\pm0.0059$ & $465\pm43$ &  & $4$ & $0.56\pm0.0318$ & $570\pm 31$ & $1.29\pm0.309$ &  &  &  &  \\
${\rm Cloud}31$ & $161.14$ & $-46.47$ & $3.0$ & $282\pm2$ & $1\pm1$ & $1$ & $0.13\pm0.0006$ & $225\pm5$ & $0.12\pm0.009$ & $4$ & $0.40\pm0.0155$ & $269\pm 39$ & $0.43\pm0.223$ & $218$ & $7.8$ & $234$ & $4.11$ \\
$\ldots$ & $\ldots$ & $\ldots$ & $\ldots$ & $\ldots$ & $\ldots$ & $\ldots$ & $\ldots$ & $\ldots$ & $\ldots$ & $\ldots$ & $\ldots$ & $\ldots$ & $\ldots$ & $\ldots$ & $\ldots$ \\
\enddata
\tablenotetext{a}{The dust cloud's series number (Col. 1), Galactic coordinates (Cols. 2 and 3) and cloud linear radius (Col. 4), the distance and the errors (Col. 5), the thickness and the errors (Col. 6), the distance level (Col. 7), the foreground fitting parameters (Col. 8 and 9) and the color excess jump in the optical (Col. 10), the times of cloud linear radius (Col. 11), the foreground fitting parameters (Col. 12 and 13) and the color excess jump (Col. 14) in the optical-ultraviolet bands, the mass and surface mass density (Col. 15 and 16), the gas-to-dust mass ratio (Col. 17), and the color excess ratio of dust clouds (Col. 18). (The Table is available in its entirety is available online)}
\end{deluxetable} 

\begin{deluxetable}{ccccccccc}
\caption{\label{table2} Comparison of our cloud distances with MBM}
\tablehead{\colhead{$Cloud_{this\ work}${\tablenotemark{a}}} & \colhead{$Cloud_{MBM}${\tablenotemark{b}}} & \colhead{$l_{this\ work}${\tablenotemark{a}}} & \colhead{$b_{this\ work}${\tablenotemark{a}}} & \colhead{$r_{this\ work}${\tablenotemark{a}}} & \colhead{$l_{MBM}${\tablenotemark{b}}} & \colhead{$b_{MBM}${\tablenotemark{b}}} & \colhead{$r_{MBM}${\tablenotemark{b}}} & \colhead{$flag${\tablenotemark{c}}} \\
\colhead{ } &\colhead{ } & \colhead{$(^{\circ})$}& \colhead{$(^{\circ})$}& \colhead{$(^{\circ})$} & \colhead{$(^{\circ})$} & \colhead{$(^{\circ})$} & \colhead{$(^{\circ})$}  &\colhead{ } \\
\colhead{(1)} & \colhead{(2)}& \colhead{(3)} & \colhead{(4)} & \colhead{(5)}  & \colhead{(6)} & \colhead{(7)}  & \colhead{(8)} & \colhead{(9)}
}
\startdata
${\rm Cloud} 11$ & ${\rm MBM} 15$ & $190.92$ & $-52.78$ & $2.09$ & $191.67$ & $-52.29$ &  & 1 \\
${\rm Cloud} 17$ & ${\rm MBM} 51$ & $72.13$ & $-49.60$ & $2.35$ & $73.31$ & $-51.53$ & $0.23$ & 0 \\
${\rm Cloud} 17$ & ${\rm MBM} 52$ & $72.13$ & $-49.60$ & $2.35$ & $74.43$ & $-51.25$ & $0.17$ & 0 \\
${\rm Cloud} 20$ & ${\rm MBM}  5$ & $144.74$ & $-50.26$ & $0.87$ & $145.97$ & $-49.07$ & $2.42$ & 1 \\
${\rm Cloud} 29$ & ${\rm MBM}  3$ & $129.84$ & $-46.50$ & $1.06$ & $131.29$ & $-45.68$ & $2.02$ & 1 \\
${\rm Cloud} 32$ & ${\rm MBM}  3$ & $134.21$ & $-45.12$ & $1.30$ & $131.29$ & $-45.68$ & $2.02$ & 0 \\
${\rm Cloud} 32$ & ${\rm MBM}  4$ & $134.21$ & $-45.12$ & $1.30$ & $133.51$ & $-45.30$ & $0.46$ & 1 \\
${\rm Cloud} 36$ & ${\rm MBM}  9$ & $157.03$ & $-45.25$ & $0.81$ & $156.53$ & $-44.72$ & $0.42$ & 1 \\
${\rm Cloud} 50$ & ${\rm MBM} 55$ & $88.11$ & $-41.25$ & $1.57$ & $89.19$ & $-40.94$ & $2.00$ & 1 \\
${\rm Cloud} 60$ & ${\rm MBM}  1$ & $110.85$ & $-40.89$ & $0.80$ & $110.19$ & $-41.23$ & $0.25$ & 1 \\
${\rm Cloud} 64$ & ${\rm MBM}  6$ & $145.70$ & $-39.30$ & $0.56$ & $145.06$ & $-39.35$ &  & 1 \\
${\rm Cloud} 67$ & ${\rm MBM}  7$ & $151.28$ & $-38.43$ & $0.61$ & $150.43$ & $-38.07$ &  & 1 \\
${\rm Cloud} 67$ & ${\rm MBM}  8$ & $151.28$ & $-38.43$ & $0.61$ & $151.75$ & $-38.67$ & $0.21$ & 1 \\
${\rm Cloud} 73$ & ${\rm MBM} 16$ & $170.97$ & $-37.53$ & $1.05$ & $170.60$ & $-37.27$ &  & 1 \\
${\rm Cloud} 74$ & ${\rm MBM} 54$ & $92.96$ & $-37.22$ & $0.75$ & $91.62$ & $-38.10$ &  & 0 \\
${\rm Cloud} 78$ & ${\rm MBM} 18$ & $189.49$ & $-35.90$ & $1.09$ & $189.10$ & $-36.02$ & $0.96$ & 1 \\
${\rm Cloud} 83$ & ${\rm MBM} 16$ & $170.91$ & $-36.02$ & $0.69$ & $170.60$ & $-37.27$ &  & 0 \\
${\rm Cloud} 86$ & ${\rm MBM} 46$ & $40.51$ & $-35.35$ & $1.28$ & $40.52$ & $-35.47$ & $0.17$ & 1 \\
${\rm Cloud} 86$ & ${\rm MBM} 47$ & $40.51$ & $-35.35$ & $1.28$ & $41.00$ & $-35.86$ & $0.17$ & 1 \\
${\rm Cloud} 86$ & ${\rm MBM} 48$ & $40.51$ & $-35.35$ & $1.28$ & $40.59$ & $-36.61$ & $0.19$ & 1 \\
${\rm Cloud} 94$ & ${\rm MBM} 53$ & $92.27$ & $-34.74$ & $1.01$ & $93.97$ & $-34.06$ &  & 0 \\
${\rm Cloud} 96$ & ${\rm MBM} 11$ & $158.70$ & $-33.07$ & $2.15$ & $157.98$ & $-35.06$ & $0.17$ & 0 \\
${\rm Cloud} 96$ & ${\rm MBM} 12$ & $158.70$ & $-33.07$ & $2.15$ & $159.35$ & $-34.32$ &  & 1 \\
${\rm Cloud} 96$ & ${\rm MBM} 14$ & $158.70$ & $-33.07$ & $2.15$ & $162.46$ & $-31.86$ & $2.38$ & 0 \\
${\rm Cloud}104$ & ${\rm MBM} 53$ & $94.17$ & $-34.17$ & $0.63$ & $93.97$ & $-34.06$ &  & 1 \\
${\rm Cloud}117$ & ${\rm MBM} 14$ & $162.27$ & $-31.70$ & $0.59$ & $162.46$ & $-31.86$ & $2.38$ & 1 \\
${\rm Cloud}132$ & ${\rm MBM} 14$ & $163.03$ & $-29.33$ & $1.11$ & $162.46$ & $-31.86$ & $2.38$ & 0 \\
${\rm Cloud}135$ & ${\rm MBM} 21$ & $209.07$ & $-29.09$ & $0.57$ & $208.42$ & $-28.39$ & $0.17$ & 1 \\
${\rm Cloud}151$ & ${\rm MBM} 49$ & $65.09$ & $-26.96$ & $0.70$ & $64.50$ & $-26.54$ &  & 1 \\
${\rm Cloud}155$ & ${\rm MBM} 56$ & $103.22$ & $-26.52$ & $0.57$ & $103.08$ & $-26.06$ & $0.50$ & 1 \\
${\rm Cloud}161$ & ${\rm MBM} 17$ & $167.43$ & $-26.19$ & $0.64$ & $167.53$ & $-26.61$ & $0.25$ & 1 \\
${\rm Cloud}163$ & ${\rm MBM} 56$ & $103.72$ & $-25.00$ & $0.96$ & $103.08$ & $-26.06$ & $0.50$ & 0 \\
${\rm Cloud}180$ & ${\rm MBM}153$ & $0.45$ & $-20.20$ & $2.24$ & $0.94$ & $-20.21$ &  & 1 \\
${\rm Cloud}180$ & ${\rm MBM}154$ & $0.45$ & $-20.20$ & $2.24$ & $1.31$ & $-20.48$ &  & 1 \\
${\rm Cloud}180$ & ${\rm MBM}155$ & $0.45$ & $-20.20$ & $2.24$ & $1.59$ & $-21.29$ &  & 1 \\
${\rm Cloud}202$ & ${\rm MBM}106$ & $177.44$ & $-20.17$ & $1.24$ & $176.33$ & $-20.78$ &  & 1 \\
${\rm Cloud}202$ & ${\rm MBM}107$ & $177.44$ & $-20.17$ & $1.24$ & $177.65$ & $-20.34$ &  & 1 \\
${\rm Cloud}202$ & ${\rm MBM}108$ & $177.44$ & $-20.17$ & $1.24$ & $178.24$ & $-20.34$ &  & 1 \\
${\rm Cloud}202$ & ${\rm MBM}109$ & $177.44$ & $-20.17$ & $1.24$ & $178.93$ & $-20.10$ &  & 1 \\
\enddata
\tablenotetext{a}{The cloud number (Col. 1), Galactic coordinates (Cols. 3 and 4), and angular radius (Col. 5) from this work}
\tablenotetext{b}{The Cloud number (Col. 2), Galactic coordinates (Cols. 6 and 7), and angular radius (Col. 8) from MBM}
\tablenotetext{c}{Cloud position match flag ( `0' and  `1' respectively represent that the main body of MBM cloud is not and is within the cloud of this work.) (Col. 9)}
\end{deluxetable}

\begin{deluxetable}{ccccccccc}
\startdata
${\rm Cloud}204$ & ${\rm MBM}101$ & $158.44$ & $-20.89$ & $0.58$ & $158.19$ & $-21.41$ &  & 1 \\
${\rm Cloud}204$ & ${\rm MBM}102$ & $158.44$ & $-20.89$ & $0.58$ & $158.56$ & $-21.15$ &  & 1 \\
${\rm Cloud}204$ & ${\rm MBM}103$ & $158.44$ & $-20.89$ & $0.58$ & $158.88$ & $-21.55$ &  & 1 \\
${\rm Cloud}204$ & ${\rm MBM}104$ & $158.44$ & $-20.89$ & $0.58$ & $158.41$ & $-20.44$ &  & 1 \\
${\rm Cloud}220$ & ${\rm MBM}151$ & $22.06$ & $20.55$ & $0.62$ & $21.53$ & $20.93$ &  & 1 \\
${\rm Cloud}227$ & ${\rm MBM}113$ & $338.04$ & $22.53$ & $1.08$ & $337.74$ & $23.04$ &  & 1 \\
${\rm Cloud}243$ & ${\rm MBM}118$ & $344.52$ & $23.86$ & $0.74$ & $344.02$ & $24.76$ &  & 1 \\
${\rm Cloud}243$ & ${\rm MBM}120$ & $344.52$ & $23.86$ & $0.74$ & $344.23$ & $24.19$ &  & 1 \\
${\rm Cloud}243$ & ${\rm MBM}121$ & $344.52$ & $23.86$ & $0.74$ & $344.78$ & $23.86$ &  & 1 \\
${\rm Cloud}243$ & ${\rm MBM}122$ & $344.52$ & $23.86$ & $0.74$ & $344.83$ & $23.92$ &  & 1 \\
${\rm Cloud}243$ & ${\rm MBM}124$ & $344.52$ & $23.86$ & $0.74$ & $343.97$ & $22.73$ &  & 0 \\
${\rm Cloud}244$ & ${\rm MBM} 23$ & $170.81$ & $25.24$ & $1.91$ & $171.84$ & $26.71$ & $0.19$ & 1 \\
${\rm Cloud}244$ & ${\rm MBM} 24$ & $170.81$ & $25.24$ & $1.91$ & $172.27$ & $26.96$ & $0.25$ & 1 \\
${\rm Cloud}252$ & ${\rm MBM}115$ & $342.65$ & $24.19$ & $0.59$ & $342.33$ & $24.15$ &  & 1 \\
${\rm Cloud}252$ & ${\rm MBM}116$ & $342.65$ & $24.19$ & $0.59$ & $342.71$ & $24.51$ &  & 1 \\
${\rm Cloud}252$ & ${\rm MBM}117$ & $342.65$ & $24.19$ & $0.59$ & $343.00$ & $24.09$ &  & 1 \\
${\rm Cloud}274$ & ${\rm MBM} 25$ & $173.46$ & $31.09$ & $0.99$ & $173.75$ & $31.48$ &  & 1 \\
${\rm Cloud}278$ & ${\rm MBM} 36$ & $5.39$ & $35.78$ & $2.14$ & $4.23$ & $35.79$ &  & 1 \\
${\rm Cloud}278$ & ${\rm MBM} 37$ & $5.39$ & $35.78$ & $2.14$ & $6.07$ & $36.76$ & $0.62$ & 1 \\
${\rm Cloud}286$ & ${\rm MBM} 39$ & $11.43$ & $36.43$ & $0.65$ & $11.41$ & $36.28$ & $0.25$ & 1\\
\enddata
\end{deluxetable}

\clearpage


\end{CJK*}
\end{document}